\documentclass[journal=jacsat,manuscript=article]{achemso}
\usepackage{chemformula} 
\usepackage[T1]{fontenc}

\usepackage{graphicx}

\usepackage[utf8]{inputenc}
\usepackage{hyperref}       
\usepackage{url}            
\usepackage{booktabs}       
\usepackage{amsfonts}       
\usepackage{nicefrac}       
\usepackage{microtype}      
\usepackage{xcolor}         
\usepackage{natbib}
\usepackage{algorithm}
\usepackage{algpseudocode}
\usepackage{multirow}
\usepackage{amsmath}
\usepackage{caption}
\usepackage{subcaption}
\usepackage{tabularx}

\newcommand{\support}{\mathrm{support}}
\newcommand{\minsup}{\mathrm{minsup}}
\newcommand{\CB}[1]{\mathrm{C{:}#1}}

\author{Masatsugu Yamada}
\affiliation[SOKENDAI]{School of Multidisciplinary Sciences, Department of Informatics, The Graduate University for Advanced Studies, SOKENDAI}
\alsoaffiliation[NII]{National Institute of Informatics, Tokyo, Japan}
\email{masatsugu-yamada@nii.ac.jp}
\author{Mahito Sugiyama}
\alsoaffiliation[SOKENDAI]{School of Multidisciplinary Sciences, Department of Informatics, The Graduate University for Advanced Studies, SOKENDAI}
\alsoaffiliation[NII]{National Institute of Informatics, Tokyo, Japan}
\email{mahito@nii.ac.jp}

\title{Molecular Graph Generation by\\ Decomposition and Reassembling}

\abbreviations{IR,NMR,UV}

\begin{document}







\begin{abstract}
Designing molecular structures with desired chemical properties is an essential task in drug discovery and material design. However, finding molecules with the optimized desired properties is still a challenging task due to combinatorial explosion of candidate space of molecules. Here we propose a novel \emph{decomposition-and-reassembling} based approach, which does not include any optimization in hidden space and our generation process is highly interpretable. Our method is a two-step procedure: In the first decomposition step, we apply frequent subgraph mining to a molecular database to collect smaller size of subgraphs as building blocks of molecules. In the second reassembling step, we search desirable building blocks guided via reinforcement learning and combine them to generate new molecules. Our experiments show that not only can our method find better molecules in terms of two standard criteria, the penalized $\log P$ and drug-likeness, but also generate drug molecules with showing the valid intermediate molecules.
\end{abstract}

%
\section{Introduction}
 Designing new molecules for drug and material with desired properties is a challenging task due to the massive number of potential drug-like molecules, which is estimated to be between $10^{23}$ to $10^{60}$~\cite{Polishchuk2013GDB17,Kirkpatrick2004chemicalspace}. Molecules are essentially represented as \emph{graphs} with node and edge attributes, while such graph structure of chemical compounds makes it difficult to generate valid molecules with desired activity or property even if you can build a Quantitative Structure-Activity Relationship (QSAR) model, which is a computational modeling method for revealing relationships between structural properties of chemical compounds and biological activities~\cite{QSAR}, by designing descriptors of chemical features specifically for virtual screening. The straightforward way of generating molecules is to solve the inverse QSAR problem through the objective function estimated from the molecular structures~\cite{Wong2009IQSAR, Miyao2016IQSAR, Churchwell2004IQSAR}. However, feature vectors extracted from molecular graphs are often highly correlated between its features, which makes it challenging to reconstruct a new molecular graph from the optimized descriptors as it requires preserving such correlation information.
 
 A number of methods have been proposed to tackle this problem of molecular generation~\cite{Gugisch2014MOLGEN, Takeda2020MolIBM, Olivecrona2017REINVENT}. Recent advanced approaches to finding of drug-candidate molecules have employed deep generative models~\cite{Rafa2018ChemVAE, kusner2017grammar, ORGAN, Jin2018JTVAE}. The basic idea of using generative models is to learn the latent representation of molecules, which enables us to reconstruct and explore molecules that satisfy target properties in the learned latent chemical space. 
 Exploration methods such as Bayesian optimization is used to search the latent chemical space~\cite{Rafa2018ChemVAE}. However, it is fundamentally difficult to reconstruct molecular graphs from the latent space and to search molecules with the desired property by extrapolation from a training dataset as a large part of the latent space represents invalid molecules.
 
 Another strategy to search desired molecules is based on a reinforcement learning. In the setting of reinforcement learning, an agent learns the optimal policy to maximize the cumulative reward, and the trained agent can take an action to generate the optimal molecules. When each molecule is represented as a string in the form of the simplified molecular-input line-entry system (SMILES)~\cite{Weininger1988SMILES}, the agent takes an action of the next character of SMILES based on the optimized policy, where recurrent neural networks (RNNs) are often used to generate strings. In the case of molecular graph generation using reinforcement learning, the agent takes an action of choosing the atom type and bond type between nodes to expand each molecule~\cite{You2018GCPN}. The state is represented as latent feature vectors by using RNNs or graph neural networks. However, both approaches of SMILES generation and node-wise molecular graph generation share the problem that the intermediate steps do not represent valid molecules, which significantly deteriorates the interpretability of resulting generated molecules.
Moreover, the property and the state radically change if the ring structure appears, and it is fundamentally difficult to treat such binary response in optimization on a continuous latent space.
 
 In this paper, we propose a novel molecular generation approach, called MOLDR (MOLecular graph Decomposition and Reassembling), which generates optimized new molecules by decomposing molecular graphs in a training dataset into subgraphs and reassembling such obtained subgraphs again in a different way. Our key insight is that chemical properties depend on the combination of subgraphs, which correspond to the functional group or the motif of molecules in the context of chemoinformatics, and that it can be optimized when appropriate substructures are included in molecules. More specifically, MOLDR is composed of a \emph{decomposition} step and a \emph{reassembling} step. In the decomposition step, we first convert each molecular graph into a tree structure to efficiently obtain subgraphs; that is, functional groups, followed by extracting frequent subgraph structures by applying a graph mining method.
 In the reassembling step, we treat the extracted subgraphs as building blocks of molecular graphs and reassemble them by searching desired blocks according to the target property using reinforcement learning. Although MOLDR can employ other optimization methods such as Monte Carlo tree search (MCTS)~\cite{MCTS_Go,MCTS_bandit}, we consistently use reinforcement learning in our study as it is known to be effective in the context of molecular generation. We empirically evaluate molecular graphs generated by our method with respect to various well-established property scores, the penalized $\log P$ and Quantitative Estimation of Drug-likeness (QED)~\cite{Bickerton2012QED} and muti-objective score of QED and Synthetic Accessibility (SA). In addition, we evaluate our method in the task of re-discovery of known drug molecules, and show that our method is competitive to the state-of-the-art molecular generation methods, with showing transition paths of generated molecules.

Our contributions are summarized as follows:
\begin{itemize}
    \item Our method MOLDR explicitly constructs new molecules by combining substructures of molecules, hence its generation process is highly interpretable.
    \item MOLDR can easily generate larger size of molecules out of distribution in a dataset by combining subgraph structures. 
    \item Molecules generated by MOLDR are superior to those by the current state-of-the-art generative models in terms of $\log P$ and QED (drug-likeness).
\end{itemize}

\section{Related works}

\citeauthor{Yang2017ChemTS} and \citeauthor{Olivecrona2017REINVENT} proposed SMILES generation approaches by RNNs and searched molecules with desired properties over SMILES representation using MCTS and policy gradient respectively. \citeauthor{Yang2021MPChemTS} also proposed the massive parallel computation of MCTS to generate and search molecules. Instead of SMILES based strategies, \citeauthor{You2018GCPN} proposed node-wise graph generation and property optimization using reinforcement learning. States of molecules are represented through graph convolutional networks, and the agent selects nodes, edges types, and the terminal to expand molecules. The policy is optimized through the Proximal Policy Optimization. These methods can generate valid molecules with desired properties at the final step. However, the generation process is a black-box by nature and it is difficult to explain why and how such molecules are obtained.

\citeauthor{Jin2018JTVAE} proposed a VAE model that generates junction trees over molecules. Nodes in a junction tree represent subgraphs extracted from a molecular dataset, and a graph neural network determines which nodes or edges are combined with each other in the junction tree. To search molecules that optimize the desirable properties, it is necessary to search two vectors, what a tree structured scaffold is and how a molecule is reconstructed within latent embedding space. In contrast, in our method, junction trees themselves are used to efficiently extract frequent substructures from a molecular dataset, and we expand and search substructures directly to achieve target scores instead of generating junction trees from VAE. 

\citeauthor{Takeda2020MolIBM} proposed to generate molecules by combining substructures that contribute to the target properties, where candidate molecules are searched by McKay’s Canonical Construction Path (MC-MCCP) algorithm~\cite{MCKAY1998306, MCKAY_Label}. \citeauthor{Jin2020multirationale} proposed the multi-objective molecule generation using interpretable substructures as rational for extracting substructures by MCTS to generate molecules by merging common substructures and graph completion. Although their approaches and our approach share the general strategy of constructing new molecules from its substructures, our method can cover a wider variety of substructures in molecular generation as we directly apply frequent subgraph mining to the entire molecular dataset, which will lead to better new molecules.
 
 Our approach of combining decomposition of molecules into subgraphs by graph mining and reassembling of subgraphs to generate new molecular graph has not been studied at sufficient depth. There is a related approach in the task of planning of chemical synthesis, which also combines subgraphs and MCTS~\cite{Segler2018}, while it is not applied to the property optimization.

\section{The Proposed Algorithm: MOLDR}
 We introduce our molecular generation algorithm MOLDR. We provide the problem setting, tree decomposition preprocessing, graph mining, and the strategy to build up molecules via reinforcement learning.

\subsection{Problem Setting}
 A \emph{graph} is a tuple $G = (V, E)$, where $V$ and $E$ denote the set of nodes and edges, respectively.
 Nodes and edges can have labels (attributes) via label functions $l_V : V \to \Sigma_V$ for nodes and $l_E : E \to \Sigma_E$ for edges with some label domain $\Sigma_V$, $\Sigma_E$, which can be any set such as $\mathbb{Z}$ and $\mathbb{R}^d$.
 We assume that each molecule is represented as a graph. If we see a graph as a molecule, $V$ is the set of atom types, and $E$ is the set of bond types. 
 For two graphs $G = (V, E)$ and $G' = (V', E')$, we say that $G'$ is a \emph{subgraph} of $G$, denoted by $G' \sqsubseteq G$, if $V' \subseteq V$ and $E' \subseteq (V' \times V') \cap E$.
 
 Let $f(G)$ be some chemical property of a graph $G$, which is usually a real-valued function, and we assume that $f$ is known beforehand and we can compute $f(G)$ for any graph $G$. For example, $f$ can be the $\log P$ of a molecular $G$.
 Given a molecular dataset, which is a collection of graphs, the problem of molecular generation is to explore a new graph $G_{\text{new}}$ that has a high $f(G_{\text{new}})$ value as long as possible.
 

\subsection{Graph Decomposition via Frequent Subgraph Mining}\label{subsec:fsm}
 Given a collection of graphs as an input molecular dataset, our idea is to apply frequent subgraph mining~\cite{Zaki2014} to the dataset, which finds all subgraphs that frequently appear in the graph dataset.
 Formally, given a graph dataset $\mathbf{D}=\{G_1, G_2, \ldots, G_n\}$ that contains $n$ graphs, the objective of frequent subgraph mining is to find all subgraphs $G$ satisfying the condition $\support(G) \geq \minsup$, where $\support(G)$ is defined as
 \begin{align*}
     \support(G) = \Big|\left\{G_i \in \mathbf{D} \mid G \sqsubseteq G_i\right\}\Big|,
 \end{align*}
 that is, the number of graphs in $\mathbf{D}$ that contains $G$ as a subgraph, and $\minsup \in \mathbb{N}$ is a frequency threshold.
 
 We use the gSpan~\cite{gSpan} algorithm, which is commonly used for the task of frequent subgraph mining.
 It enumerates subgraphs in a depth first manner.
 In gSpan, each graph is represented as the DFS code, which is constructed from a search tree based on a lexicographic order and enables us to efficiently check duplication of enumerated graphs. More precisely, 
 for each explored graph during the enumeration, it checks whether or not its DFS code is canonical.
 After completion of gSpan, we check every enumerated subgraph and keep only subgraphs whose target property score is already higher than some threshold, which is determined beforehand, to efficiently reassemble them to construct new graphs in the next reassembling step.
 
 Molecules are firstly converted into molecular graphs, where each node represents an atom type and each edge represents a bond type. However, if we directly apply gSpan to such molecular graphs, it gives a lot of invalid subgraphs in terms of molecules as building blocks for molecular generation. This is because gSpan does not know the chemical context and simply enumerates frequent subgraphs, hence, for example, the ring structure will be truncated by gSpan, while such truncated subgraphs are invalid and unnecessary for the reassembling step.
 
 To circumvent this problem, we apply \emph{tree decomposition} to molecular graphs as preprocessing before applying gSpan, and convert them into molecular junction trees.
 A tree decomposition maps a graph $G=(V, E)$ into a \emph{junction tree} $\mathcal{T}=(\mathcal{V}, \mathcal{E})$, where $\mathcal{V}=\{C_1, \dots, C_n\}$ is a collection of subsets of $V$; that is, each $v_i \subseteq V$, and $\mathcal{E}$ is a set of edges between elements of $\mathcal{V}$.
 A junction tree satisfies the following properties:
\begin{enumerate}
 \item The union of all sets $C_1$, $\dots$, $C_n$ equals to $V$; that is, $\bigcup_i C_i = V$.
 \item For every edge $(u, v) \in E$, there exists $C_i \in \mathcal{V}$ such that $u \in C_i$ and $v \in C_i$. 
 \item If $C_k$ is on a path from $C_i$ to $C_j$ in $\mathcal{T}$, $V_i \cap V_j \subseteq{V_k}$. 
 \end{enumerate}
 By converting a graph into its corresponding junction tree, by definition, each cycle will be gathered as a single node and all cycles will be eliminated. Therefore, if we apply gSpan to not the original graphs but the converted junction trees, we can avoid enumerating invalid subgraphs in which the ring structure of a molecule, represented as a cycle on a graph, is truncated.
 In addition, gSpan on junction trees can dramatically reduce the number of frequent subgraphs. This is also an advantage of using junction trees in the decomposition step for molecular generation.
 
 The edge label information and the node label information in each clique are lost in a junction tree, hence we need to restore them after frequent subgraph mining.
 To achieve this task, we use a subgraph matching algorithm that matches between original graphs and obtained trees.
 We use the indexed based subgraph matching algorithm with general symmetries (ISMAGS)~\cite{ISMAGS}. 
 Since the size of each molecule is usually not so large and the number of nodes is mostly around 20$\sim$30 in the task of molecular generation, this restoring process is not computationally expensive.

\subsection{Graph Reassembling from Frequent Subgraphs}
 
Now we generate new molecules by reassembling frequent subgraphs obtained by the previous graph decomposition step. In contrast to our approach using subgraphs as building blocks, existing approaches are based on either text generation or node-wise graph generation.
In the text generation approach~\cite{SMILES_RNN} based on SMILES, an algorithm picks up a particular character which denotes chemical state, such as the atom (\ch{C}, \ch{N}, \ch{O}, \ch{F}, ...), the bond type (=, $\equiv$), or the branched symbols, from the set of character types occurred in a training dataset to generate and expand molecules. 
In the node-wise graph generation~\cite{You2018GCPN, Li2018DGMG}, an algorithm selects a node (atom symbol) and the edge type between source and target atoms from the candidate set of atom and edge types. Our method can be more powerful and efficient as we directly combine subgraphs that already have desirable properties as building blocks in molecular generation.

\begin{figure}
    \centering
    \includegraphics[width=0.5\textwidth]{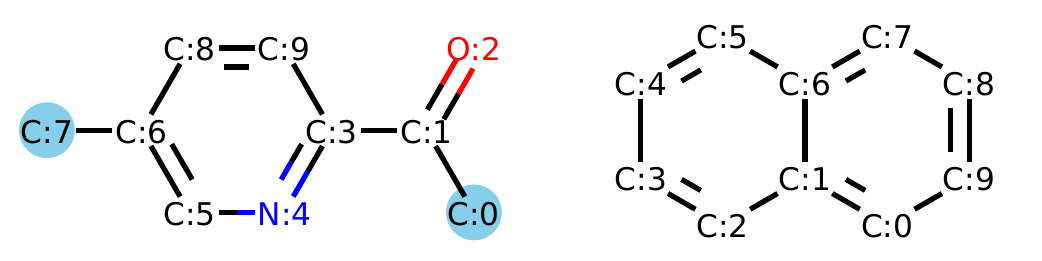} 
    \caption{An example of molecular graphs $G$ and $G'$. Atoms of C:0 and C:7 are candidates to merge.}
    \label{fig:reassemble_molecules}
\end{figure}

To assemble molecular subgraphs, we pick up two graphs $G_t$ and $G_t'$ from building blocks and combine them to generate a new graph $G_{t+1}$, where $t$ is the number of building up steps of molecules. As an example of such molecules, \textit{2-Acetyl-5-methylpyridine} and \textit{naphthalene} is shown in Figure~\ref{fig:reassemble_molecules}. Let us assume that $G_{t} = \left(V(G_t), E(G_t)\right)$ with $V(G_t) = \{v_1, \ldots, v_n\}$ and $G'_{t} = (V(G'_t), E(G'_t))$ with $V(G_t') = \{u_1, \dots, u_{n'}\}$. In the reassembling procedure with nodes, we select single nodes $v_i \in V(G_t)$ and $u_j \in V(G_t')$ such that they have the same node labels: $l_v(v_i) = l_v(u_j)$. We overlay these two nodes as $v_{t+1}$; that is, $V(G_{t+1}) = V(G_t) \cup V(G_t') \setminus \{v_i, u_j\} \cup \{v_{t + 1}\}$ for a newly constructed graph $G_{t+1}$. All edges in $G_t$ and $G_t'$ are preserved in $G_{t+1}$, where if there is an edge $(v_i, v_k)$ or $(u_j, u_l)$, it is replaced with $(v_{t+1}, v_k)$ or $(v_{t+1}, u_l)$. In the reassembling with edges, we select edges from rings, and overlay them in the same manner as the assembling with edges.

\begin{figure}
    \centering
    \includegraphics[width=\textwidth]{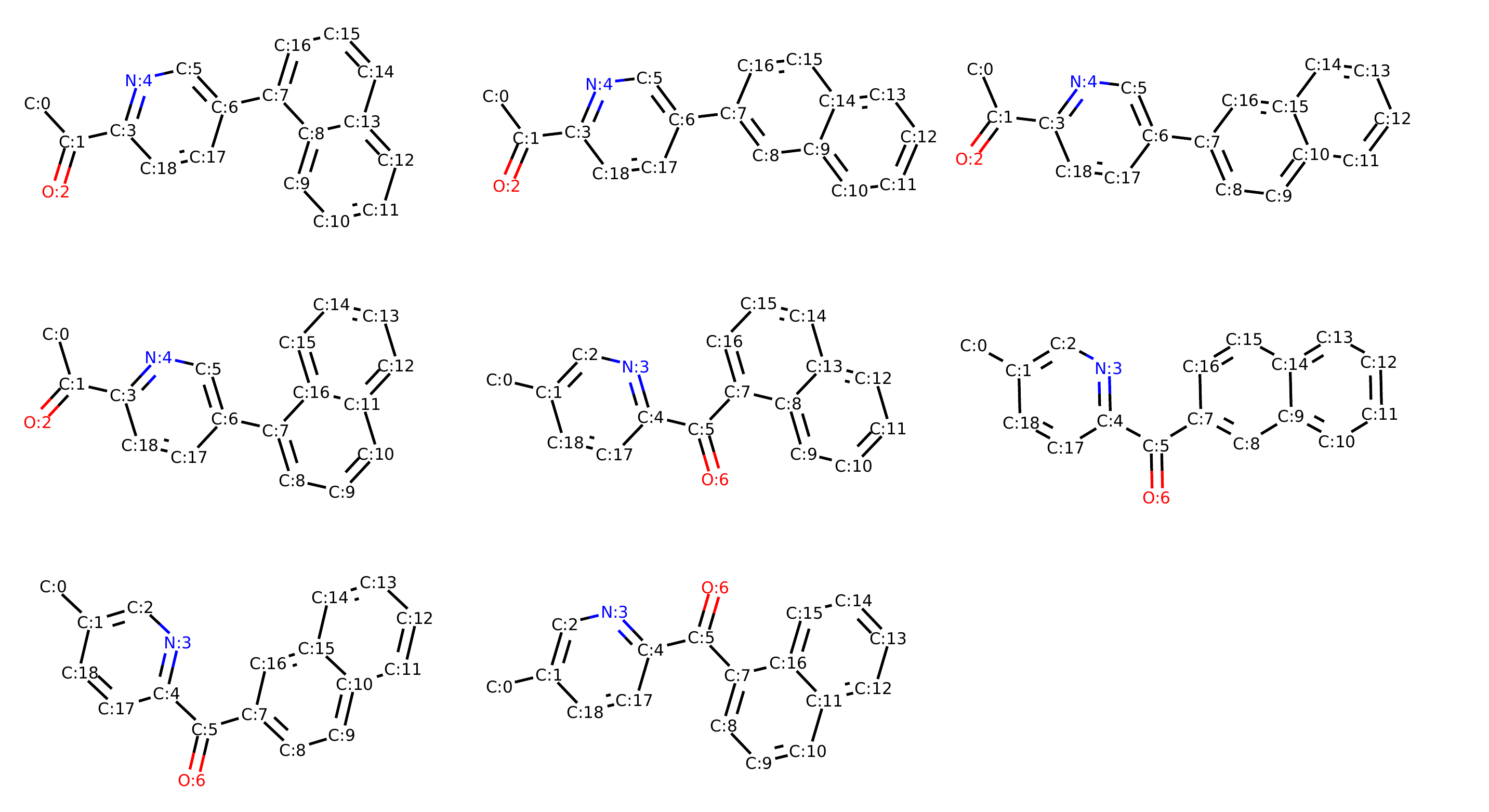} 
    \caption{Reassembling two molecules with \textbf{nodes} in Figure~\ref{fig:reassemble_molecules}. It shows merging with nodes labeled as \ch{C}. In this examples, reassembled molecules are sanitized to be valid molecules.}
    \label{fig:reassemble_node}
\end{figure}

\begin{figure}
    \centering
    \includegraphics[width=\textwidth]{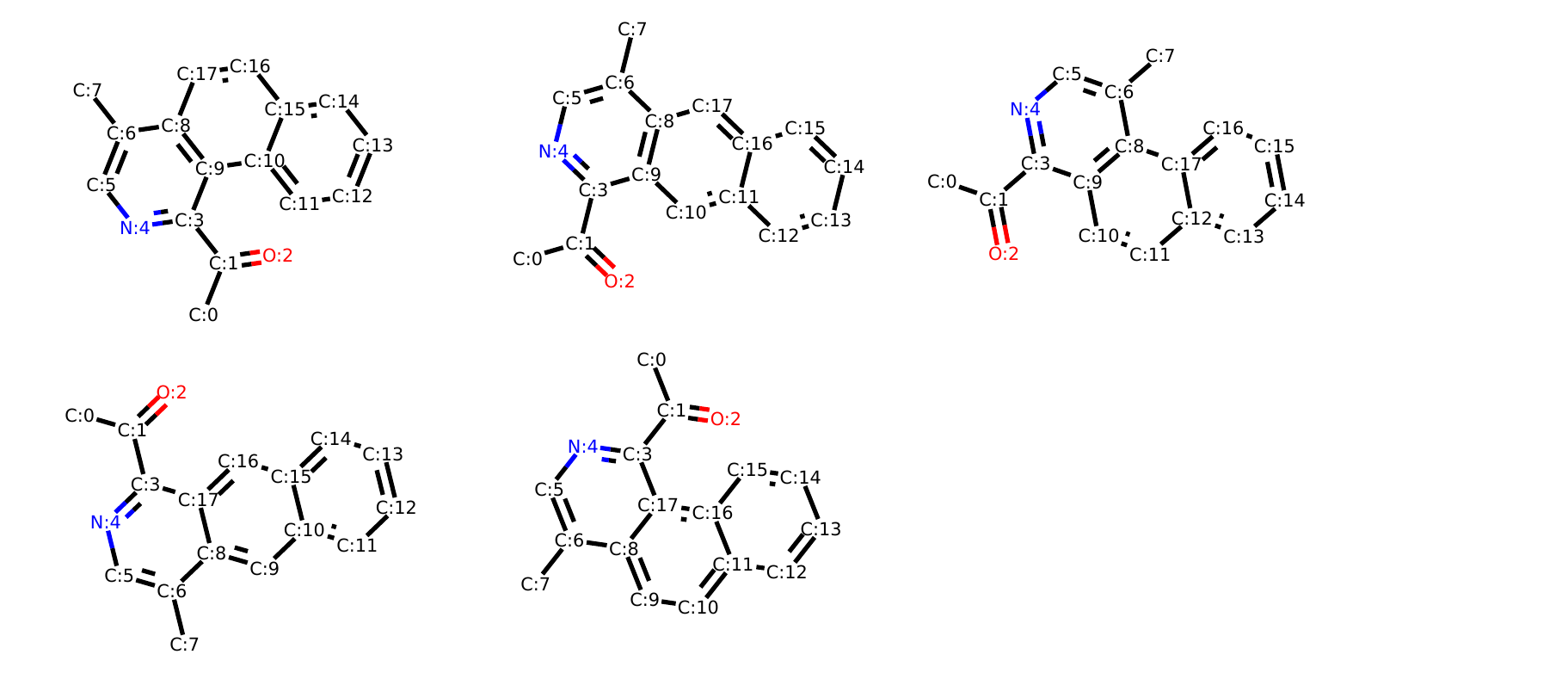}
    \caption{Reassembling two molecules with \textbf{edges} in Figure~\ref{fig:reassemble_molecules}. Molecules were merged between edges in rings.}
    \label{fig:reassemble_edge}
 \end{figure}

This assembling is similar to reconstructing a graph from a junction tree; that is, nodes of a clique in a junction tree have intersected nodes that are connected with each other between subgraphs. Assembling two graphs is equivalent to choose the intersection of nodes or edges. Figure~\ref{fig:reassemble_node} and ~\ref{fig:reassemble_edge} show the process of reassembling the two molecular graphs with nodes or edges, respectively. The candidate set of node label \ch{C} for merging is $\{\CB{0}: \{\CB{0}, \CB{2}, \CB{3}, \CB{4}, \CB{5}, \CB{7}, \CB{8}, \CB{9}\}, \CB{7}:  \{\CB{0}, \CB{2}, \CB{3}, \CB{4}, \CB{5}, \CB{7}, \CB{8}, \CB{9} \}$, where indices of nodes correspond to numbers in the illustration in Figure~\ref{fig:reassemble_molecules}. We do not include internal nodes such as $\CB{1}, \CB{6}$ as the resulting graph will be an invalid molecule nor include duplicated structures. In the reassembling with edges, the candidate set in the same node label (\ch{C}, \ch{C}) is $\{(\CB{8}, \CB{9}): \{(\CB{0}, \CB{9}), (\CB{2}, \CB{3}), (\CB{3}, \CB{4}), (\CB{4}, \CB{5}), (\CB{8}, \CB{9})\}\}$, resulting in 5 new graphs as K{\'e}kule structures, and  $(\CB{0}, \CB{9})$ and $(\CB{7}, \CB{8})$ are also the same due to symmetry structure.

 The computational cost of combining two graphs depends on the number of nodes and the number of edges in rings. In the worst case, where we need to consider all combinations of nodes and edge of two graphs $G = (V, E)$ and $G' = (V', E')$, the complexity becomes $\mathcal{O}(|V||V'| + |E||E'|)$. However, this can be usually reduced in practice by considering the symmetrical structure of a graph and some type of restrictions of chemical valency of an element in the case of molecules. We always check such conditions whenever a new molecule is generated and remove it if does not satisfy such conditions. Therefore molecules generated by our method is always valid.
 
\subsection{Finding Candidate Subgraphs by Reinforcement Learning}
To efficiently find subgraphs that will lead to desirable molecules when assembled in graph generation, we use reinforcement learning. Other potential choice of a searching method is Monte Carlo tree search (MCTS), which is a search method that combines tree search with random sampling ~\cite{MCTS_survey}. MCTS has been applied to a number of tasks when the search space is massive and achieved huge success in various fields such as the game of Go~\cite{Silver2016} and the planning of chemical syntheses~\cite{Segler2018}. However, the parallel computation of MCTS is difficult to implement, so we exploit the reinforcement learning and proximal policy optimization (PPO) that can be used for continuous control. In reinforcement learning, the agent takes action $a$ based on a policy $\pi$, which is often represented as neural networks. The policy network returns the probability of each action $a \in \mathcal{A}$ and the state value function $V_\pi$ based on the state $s \in \mathcal{S}$. In other words, at a time step $t$, the action $a_t$ is sampled with probability $V_{\pi_{\theta}}(s_t) = \pi_{\theta}(s_t | a_t)$. The agent is trained so that the expected cumulative reward $\mathbb{E}[\sum_{t=0}^{\infty} r_t]$ is maximized while interacting the environment. 

PPO is based on the trust region policy optimization (TRPO) method~\cite{Schulman2015TRPO} to prevent the high variance in learning with policy gradient~\cite{Schulman2017ppo}. The main objective is to optimize the parameter $\theta$ of a policy network $\pi$ through the loss function $L$ as following:
\begin{align*}
    L^{\text{CLIP}}(\theta) &= \hat{\mathbb{E}}_t \left[\min \right(x_t(\theta)\hat{A}_t, \text{clip}(x_t(\theta), 1 - \epsilon, 1 + \epsilon)\hat{A}_t \left) \right],\ \text{where}\\
    x_t (\theta) &= \frac{\pi_{\theta} (a_t \vert s_t)}{\pi_{\theta_{\text{old}}}(a_t \vert s_t)}, \\
    \hat{A}_t &=  \sum_{l=0}^{\infty} \gamma^{l} r_{t+l} - SV_{\pi_{\theta}} (s_t),  
\end{align*}
where $\epsilon$ is a hyperparameter and $\gamma$ is a discount factor, $SV_{\pi}$ is the state-value function denoting the expected return computed from the policy network. The function $\text{clip}(\cdot, 1-\epsilon, 1+\epsilon)$ clips the value of the first argument within the range from $1-\epsilon$ to $1+\epsilon$. The first term inside the $\min$ is conservative policy iteration~\cite{Kakade2002CPI}. The second term modifies the surrogate objective by clipping the probability ratio, which removes the incentive for moving $r_t$ outside of the interval $[1-\epsilon, 1+\epsilon]$~\cite{Schulman2017ppo}. 

 \textbf{States space}. A graph $G_t$ is mapped into $F$-dimensional continuous space in the form of a node feature matrix $H_t \in \mathbf{R}^{|V| \times F}$ converted by Mol2vec~\cite{Jaeger2018mol2vec}, which is based on word2vec~\cite{Mikolov2013word2vec} and designed for molecular substructures obtained through Morgan fingerprints. Each computed node feature in a graph is reduced via the sum function over nodes into a single $F$-dimensional vector $s_t \in \mathbf{R}^{F}$, which incorporates graph topological information.
 
 \textbf{Action space}. Action space $\mathcal{A}$ is equivalent to the set of subgraphs enumerated from subgraph mining algorithm. Each action $\mathcal{a} \in \mathcal{A}$ is sampled according to softmax of the likelihood based on the state value function $SV(s_t)$ and the expected cumulative reward.
 
 \textbf{Rewards}. To generate molecules with target properties, we assume that target properties satisfies the additive compositionality of subgraphs. Therefore, if some subgraph structure is not related to the target property, it is not selected as a building block. This assumption can be simply represented in the following strategy: If the difference of rewards at a time step $t$ is negative, we stop the trial. The reward $r_t$ is computed by a scoring function $f(G_t)$ at time step $t$. If the reward is below a certain threshold of the score, further search will be stopped. 
\begin{figure}
    \centering
    \includegraphics[width=\textwidth]{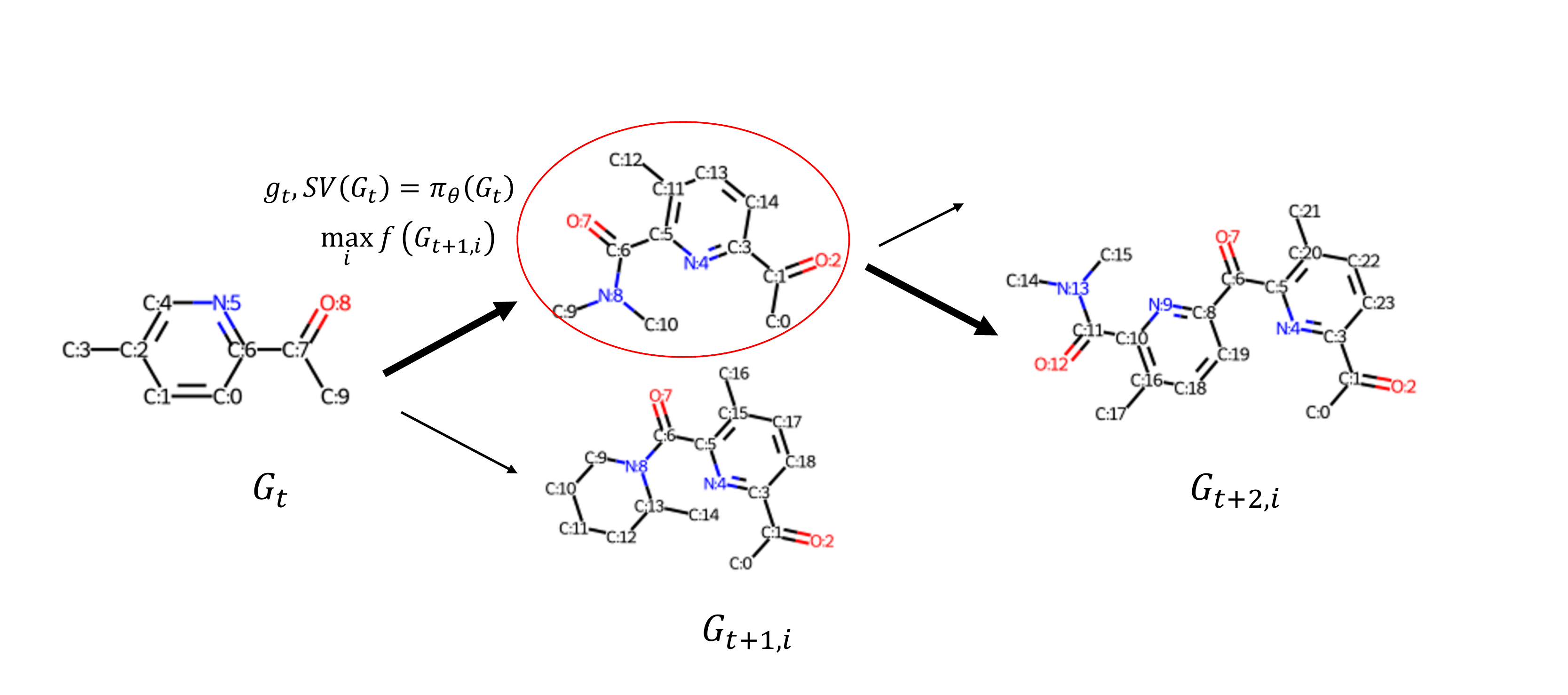}
    \caption{The diagram of molecular generation in reassembling step. Mined subgraph structures are selected based on the policy network $\pi_{\theta}$. After reassembling molecules, a new graph is selected based on the highest score of a target property computed from reward function $r$.}
    \label{fig:moldr}
\end{figure}

\section{Experiments}

We empirically examine the effectiveness of our proposed method MOLDR compared to the state-of-the-art molecular generation methods. In particular, first we examine the standard criteria, the penalized $\log P$ and the drug-likeness score QED, of generated molecules. In addition, we also examine the multi-objective score of QED and SA. Furthermore, we benchmark the rediscovery molecules using GuacaMol benchmark dataset~\cite{Brown2019GuacaMol}. 

All methods are implemented in Python 3.7.6. We used the gSpan library\footnote{\url{https://github.com/betterenvi/gSpan}} to obtain building blocks. All neural networks and reinforcement learning are implemented in RLlib~\cite{Liang2018RLlib} and PyTorch~\cite{PyTorch}. All experiments were conducted on Ubuntu 18.04.5LTS with 40 cores of 2.2 GHz Intel Xeon CPU E5-2698 v4, 256GB of memory, and 32GB Nvidia Tesla V100. 

\textbf{Dataset.} We use the ZINC molecular dataset and GuacaMol dataset with about 1.5m molecules preprocessed with the ChEMBL dataset. ZINC dataset is a freely available drug-like molecular database~\cite{Irwin2012ZINC}. There are 249,456 molecules in total, and the maximum numbers of nodes and edges are 38 and 45, respectively. The number of node labels is 8: \{\ch{B}, \ch{C}, \ch{F}, \ch{I}, \ch{N}, \ch{O}, \ch{P}, \ch{S}\}. The number of molecules on the GuacaMol dataset is 1,591,378 in total, and the maximum numbers of nodes and edges are 88 and 87. The number of node labels is 12: \{\ch{B}, \ch{Br}, \ch{C}, \ch{Cl}, \ch{F}, \ch{I}, \ch{N}, \ch{O}, \ch{P}, \ch{S}, \ch{Se}, \ch{Si}\}. All molecules are prepossessed by RDKit ~\cite{RDKIT} so that they are treated as graphs. 

Before applying MOLDR, we converted molecules in the molecular dataset into \textit{junction trees}. As a result, on the ZINC dataset, the number of cliques is 784, which are used as the node label of junction trees. The maximum number of nodes and edges in junction trees become 31 and 30, respectively. On the GuacaMol dataset, the number of cliques is 5,106 and the maximum number of nodes and edges in junction trees becomes 88 and 87, respectively. 
 
 \textbf{Experimental setting.} The gSpan algorithm was applied to converted junction trees with the minimum support of 10,000, 5,000, 1,000, and 100, where we enumerated molecules with more than 7 nodes. To see the effectiveness of our junction tree-based enumeration, we also applied gSpan to the original ZINC dataset without junction tree conversion. On the GuacaMol dataset, we applied gSpan with minimum support of 10,000.
 
 In the molecular reassembling step in MOLDR, we use building blocks extracted under the condition of the minimum support of 1,000. In reinforcement learning, we use the policy network with 3-layer MLPs (256, 128, 128 hidden units) to take actions (to choose building blocks and compute state value function), and the activation function is the ReLu function. The generalized advantage estimate (GAE) parameters are set as $\lambda=1.0$ and $\gamma=0.99$. The optimizer is stochastic gradient descent, where the mini-batch size is 128 and the learning rate is $5.0 \times 10^{-5}$. The terminal condition is when the maximum number of nodes exceeds 100, or the previous reward exceeds the current reward or threshold. 

\textbf{Target properties.}

As target chemical properties, we employ scores of the \emph{penalized $\log P$}~\cite{kusner2017grammar} and Quantitative Estimation of Drug likeness (QED)~\cite{Bickerton2012QED}. These values are widely used as a benchmark for the task of a molecular generation. The penalized $\log P$ is a logarithm of the octanol-water partition coefficient with restrictions on the ring size and synthetic accessibility (SA)~\cite{Ertl2009SA}. SA score is defined as follows:
\begin{equation*}
    \text{SA} = \text{Fragment Score}  - \text{Complex Penalty}. \\
\end{equation*}
The Fragment Score was introduced to capture the ``historical synthetic knowledge'' by analyzing common structural features in a large number of already synthesized molecules\cite{Ertl2009SA}. Complex Penalty is computed from summation of each term as follows:
\begin{align*}
    \text{Ring complexity} &= \log(\text{nRingBridgeAtoms} + 1) + \log(\text{nSprioAtoms} - 1), \\
    \text{Stereo complexity} &= \log(\text{nStereoCenters} + 1),  \\
    \text{Macro Cycle Penalty} &= \log (\text{nMacroCycles} + 1),  \\
    \text{Size penalty} &= \text{nAtoms}^{1.005} - \text{nAtoms},
\end{align*}
where ``n'' denotes the number. We used the penalized $\log P$ normalized with the ZINC250k dataset to compare the same setting with other methods, thus direct comparison of scores is fair. QED is the score representing the drug-like nature of molecular structures. QED represents the function of weighted chemical properties: 
\begin{align*}
    \text{QED} = \exp \left( \frac{\sum w_i \log d_i}{\sum w} \right),
\end{align*}
where $w$ is the weight of a molecule and each $d_i$ is one of the following chemical properties: molecular weight (MW), octanol-water partition coefficient (ALOGP), number of hydrogen bond donors (HBD), number of hydrogen bond acceptors (HBA), molecular polar surface area (PSA), number of rotatable bonds (ROTB), the number of aromatic rings (AROM), or number of structural alerts (ALERTS). Thus optimizing QED implies generating the molecules subject to these parameters. For guiding target values of property, we optimize molecule with $\log P=8.0$. For multi-objective benchmark, we generate molecules such that QED is higher and SA is smaller (easy to synthesize). We choose the objective function proposed by ~\citeauthor{Tan2022DRlinker} defined as follows: 
\begin{equation*}
    f(G) = \max\bigl(\,\text{QED}(G) - 0.1 \text{SA}(G)\,\bigr).
\end{equation*}



\textbf{Distribution benchmarks}.

To investigate whether MOLDR can generate diverse molecules or not, we use the GuacaMol benchmark dataset. The proposed scores are listed as follows: 
\begin{itemize}
    \item Validity: whether the generated molecules are actually valid computed in RDKit.
    \item Uniqueness: the ratio of molecules that are not duplicated to generate molecules. 
    \item Novelty: the ratio of molecules that are not duplicated to original dataset.
    \item Kullback-Leibler (KD) divergence: measures how well a probability distribution Q approximates another distribution P: $D_{KL}=\sum_i P(i) \log \frac{P(i)}{Q(i)}$. The probability is calculated from physiochemical descriptors for the training set and generated set. 
    \item Fréchet ChemNet Distance (FCD). \citeauthor{Preuer2018fcd} introduced the Fréchet ChemNet Distance as a measure of how close distributions of generated data are to the distribution of molecules in the training set. Low FCD values characterize similar molecule distributions
\end{itemize}

\textbf{Rediscovery molecules}.
We examined whether or not MOLDR can reconstruct target molecules such as drugs, and can generate molecules exceeding some threshold of similarity between molecules, not just generating molecules with chemical properties. In such cases, we chose Celecoxib, Troglitazone and Thiothixene as rediscovery benchmarks, Aripiprazole as similarity benchmark, and Ranolazine and Osimertinib as MPO benchmarks. 

\subsection{Results and Discussion}
\begin{figure}
    \centering
    \includegraphics[width=\linewidth]{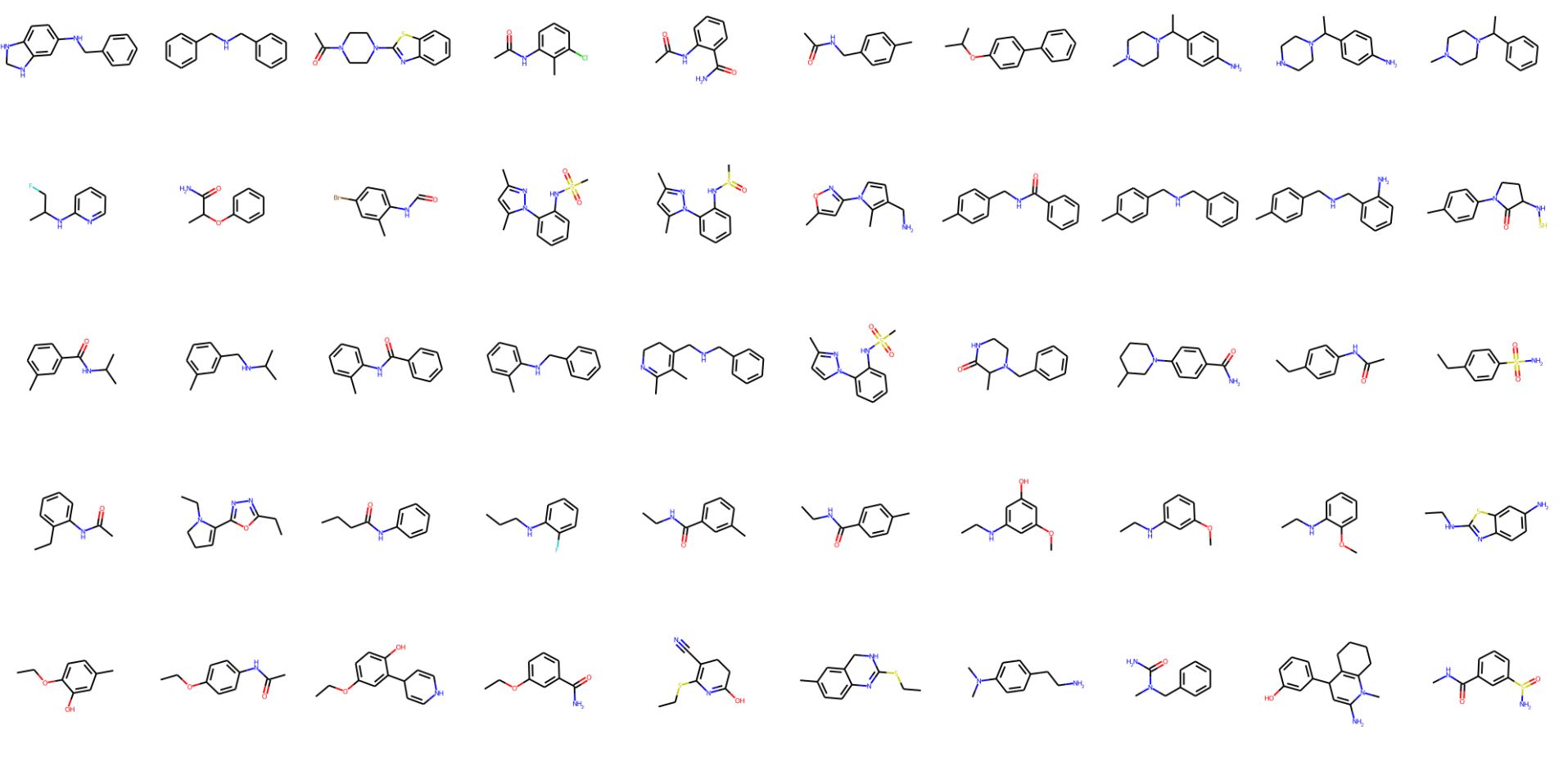} 
    \caption{Examples of extracted substructures sorted by the score of QED. ZINC 250k molecules are decomposed into junction trees, gSpan enumerates frequent subtrees, and are reconstructed into molecules by ISMAGS. These substructures become building blocks for molecular reassembling. 
    }
    \label{fig:building_block}
\end{figure}

Table~\ref{table:gspan} shows results of applying gSpan to the ZINC database with varying the minimum support. We compare the number of obtained subgraphs and calculation time with or without molecular junction trees. We can see that enumeration based on junction trees is much faster than directly applying gSpan to molecular graphs. This result means that our junction tree-based enumeration is effective in the real-world ZINC database. 
Note that we can fully recover original subgraphs from mined trees using subgraph matching as we have  discussed, thus there is no information loss in the junction tree-based enumeration and it can be viewed as loss-less compression of subgraphs.
When the minimum support is 100, we could find a large number of subgraphs (23,616 subgraphs), and it is expected that we have collected enough substructures. Therefore we stop decreasing the minimum support.
Figure~\ref{fig:building_block} shows examples of building blocks of substructures extracted from the ZINC 250k filter by the score of QED > 0.7. The obtained substructures are frequent subgraphs with minimum support of 100. These structures are used as building blocks for molecular graph reassembling. 

\begin{table}[b]
  \caption{Comparison of frequent subgraph enumeration with or without junction trees.}
  \label{table:gspan}
  \centering
  \begin{tabular}{rcc}
    \toprule
    $\minsup$ & Number of mined trees & Number of mined graphs  \\
    \midrule
    100,000 & 0 & 23 (1334 sec) \\
    10,000  & 8 (164.42 sec) & 4040 (106.5 min) \\
    5,000   & 39 (216.21 sec) & --- \\
    1,000   & 910 (342.20 sec) & ---  \\
    100     & 23,616 (775.23 sec) & --- \\
    \bottomrule
    \end{tabular}
     \\ ``---'' means that computation did not stop in 2 hours
\end{table}

\begin{table}[b]
  \caption{Comparison of the top 3 property scores of generated molecules. Scores for ORGAN, JT-VAE, and GCPN are from~\cite{You2018GCPN}, and ~\cite{Chence2020graphaf}. on the ZINC dataset.}
  \label{table:mol_gen}
  \centering
  \begin{tabular}{ ccccccccc } 
  \toprule
  \multirow{2}{*}{Method} & \multicolumn{3}{c}{Penalized $\log P$} & \multicolumn{3}{c}{QED}  \\ 
  \cline{2-9}
   & 1st & 2nd & 3rd & Validity & 1st & 2nd & 3rd & Validity\\
  \midrule
  ZINC    & 4.52  & 4.30  & 4.23  &  100.0\% & 0.948 & 0.948 & 0.948 & 100.0\% \\
  ORGAN   & 3.63  & 3.49  & 3.44  &  0.4 \%  & 0.838 & 0.814 & 0.814 & 2.2\%  \\
  JT-VAE  & 5.30  & 4.93  & 4.49  &  100.0\% & 0.925 & 0.911 & 0.910 & 100.0\% \\
  GCPN    & 7.98  & 7.85  & 7.80  &  100.0\% & \textbf{0.948} & 0.947 & 0.946 & 100.0\% \\
  GraphAF & 12.23 & 11.29 & 11.05 &  100.0\% & \textbf{0.948} & \textbf{0.948} & \textbf{0.947} & 100.0\% \\
  \midrule
  MOLDR & \textbf{12.46} & \textbf{12.20} & \textbf{12.04} & \textbf{100.0\%} & \textbf{0.948} & \textbf{0.948} & \textbf{0.947} & \textbf{100.0\%}  \\
  \bottomrule
  \end{tabular}
\end{table}

\begin{figure}
    \centering
    \includegraphics[width=\textwidth]{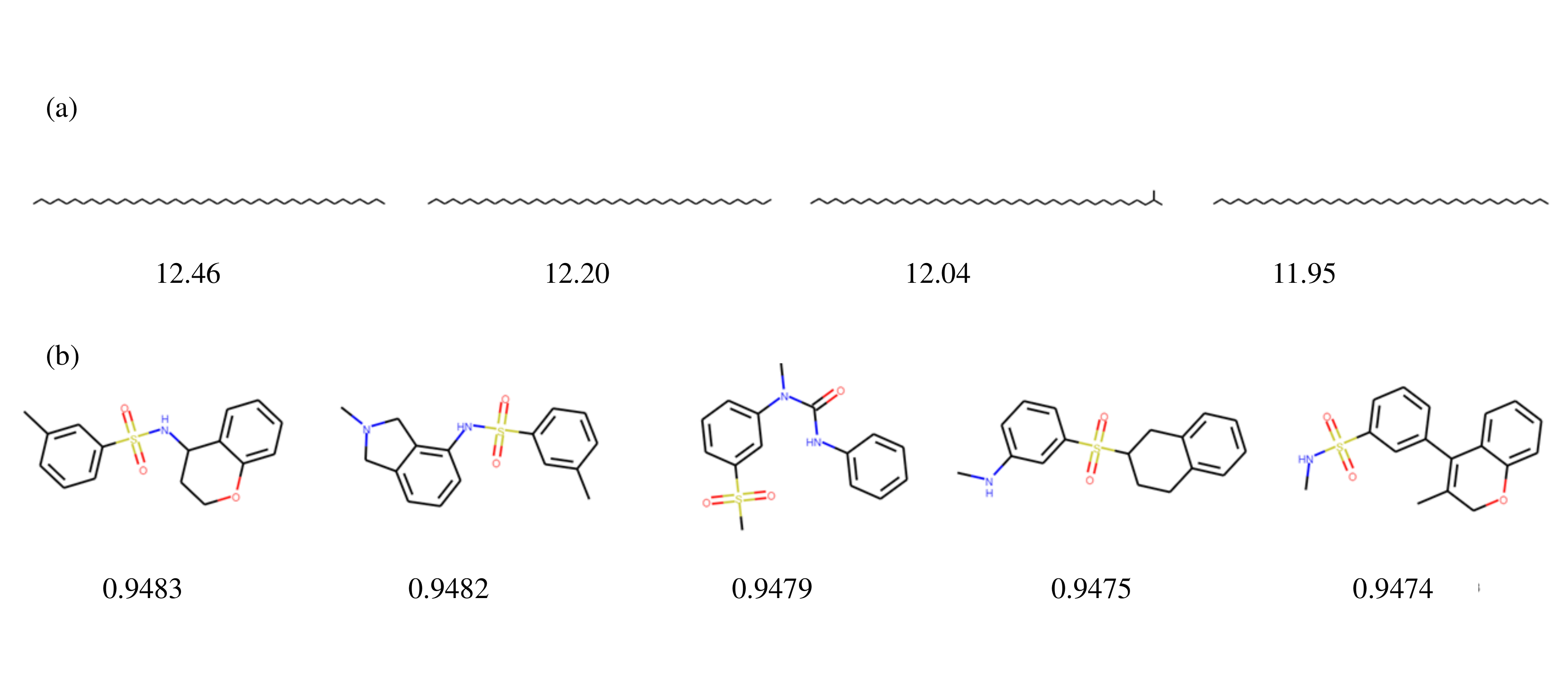} 
    \caption{Generated molecules based on ZINC dataset by MOLDR with penalized $\log P$ and QED scores.}
    \label{fig:logP_QED_ZINC}
\end{figure}

Table~\ref{table:mol_gen} shows the top 3 generated molecules according to property scores of the penalized $\log P$ or QED. Scores of other methods come from literature~\cite{You2018GCPN, Chence2020graphaf}. MOLDR is similar to the technique of JT-VAE as both methods use junction trees, while MOLDR outperforms both scores. The $\log P$ is related to the lipophilicity and hydrophilicity of a molecule. Hence, if nodes in a generated molecule have many carbon (\ch{C}) and less imide (\ch{=NH}) or hydroxyl groups (\ch{OH}), the resulting $\log P$ becomes high. It means that the larger the number of the atom \ch{C} is, the higher the $\log P$ value is. At the same time, we show penalized $\log P$ scores in which the ring size and synthetic accessibility are penalized in Table~\ref{table:mol_gen}. In the case of penalized $\log P$ optimization, an approach of greedy search such as selecting only \ch{C} can be enough to maximize the score because the calculation of the $\log P$ score consists of additive compositionality. MOLDR can train such a strategy by optimizing the penalized $\log P$ (the molecule with the top score has only \ch{C} (C43)) as shown in Figure~\ref{fig:logP_QED_ZINC} (a). The QED score is empirically derived from the combination of various chemical properties and chemical structures. Hence it is not straightforward to maximize QED, unlike the case of $\log P$.
Nevertheless, MOLDR outperforms the score of JT-VAE and top-1 and -2 molecules generated by GCPN. To increase the QED score, generated molecules need to follow the strict restriction of structures. Figure~\ref{fig:logP_QED_ZINC} (b) illustrates examples of generated molecules with optimization of QED by MOLDR. In penalized $\log P$ optimization, when the molecular size becomes larger and molecules include a large number of \ch{C}, the resulting $\log P$ increases. In QED optimization, the size of the molecule is smaller than the case of $\log P$ optimization, and they have subgraphs that contribute to the QED. We remove the similar graphs with the highest score in Figure~\ref{fig:logP_QED_ZINC} to show the variety of generated molecules by our method.

\begin{figure}
    \centering
    \includegraphics[width=\linewidth]{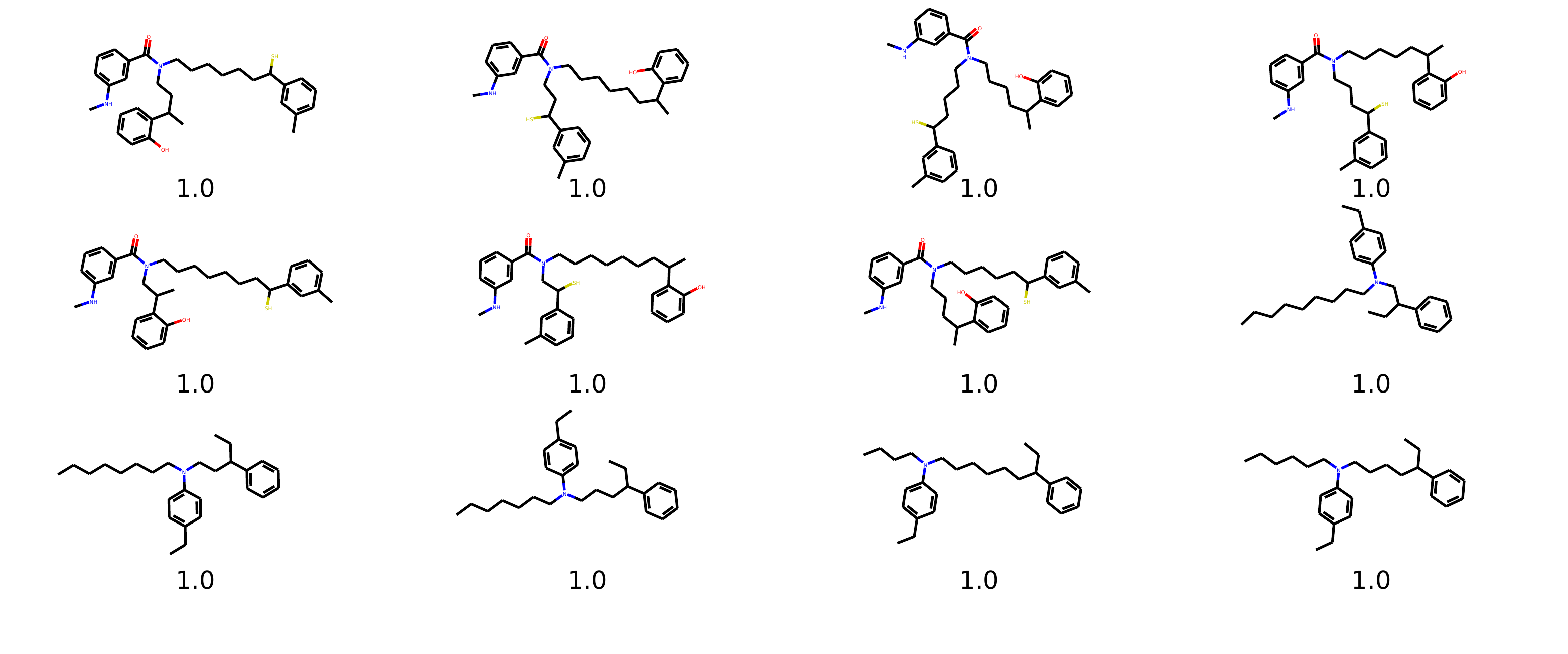}
    \caption{Generated molecules with $\log P$ = 8.0}
    \label{fig:logP}
\end{figure}

Moreover, MOLDR is flexible in the sense that it can generate molecules with not only maximizing the target value like $\log P$ but controlling it to be a specific value. As an example, we show molecules generated by MOLDR with specifying the target value $\log P=8.0$ in Figure ~\ref{fig:logP}.

\begin{figure}
\centering
    \begin{subfigure}{0.48\textwidth}
        \centering
        \includegraphics[width=\linewidth]{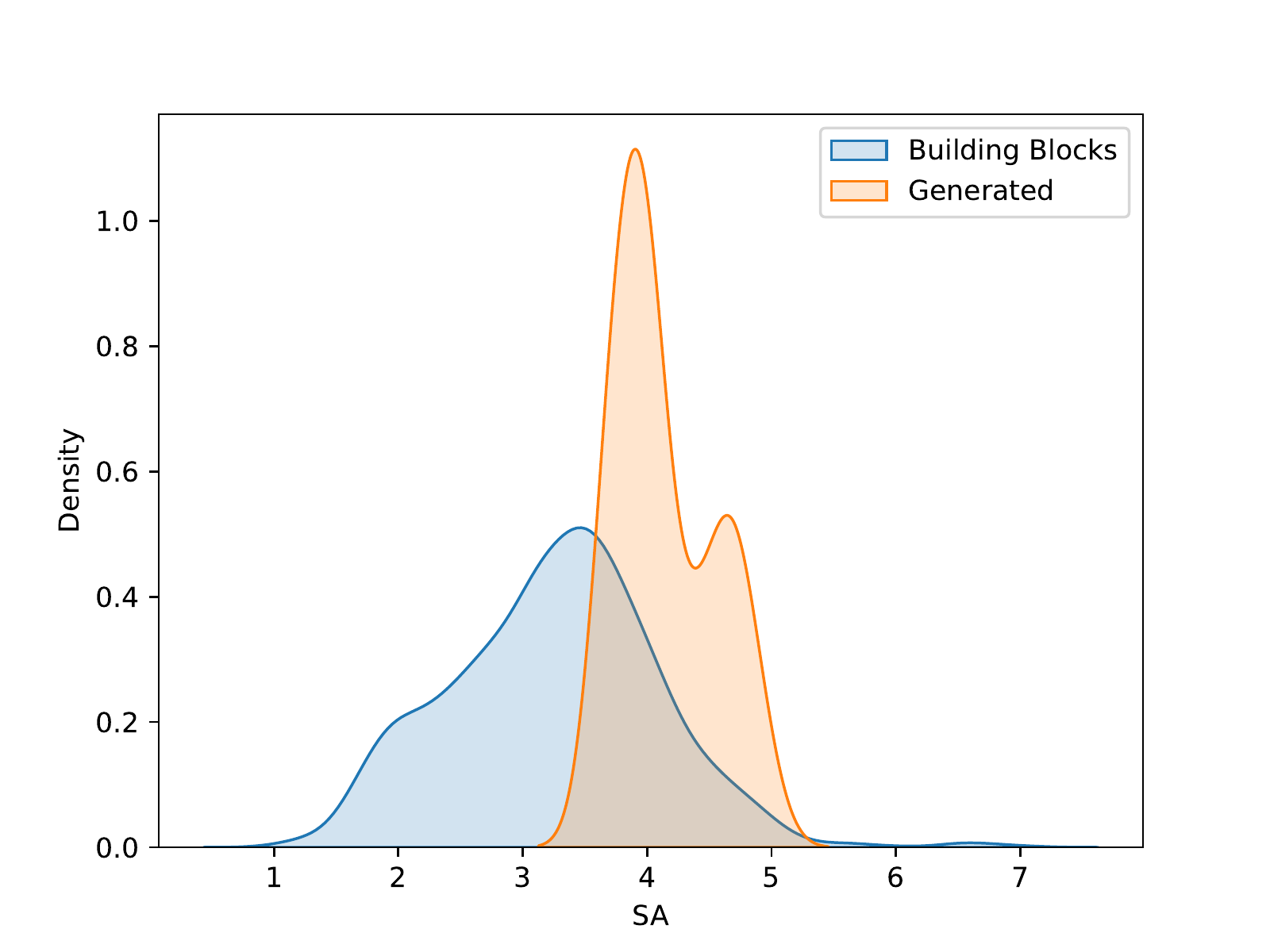}
        \caption{SA distribution of building blocks and generated molecules when QED is optimized.}
        \label{fig:dist_sa}        
    \end{subfigure}
    \hspace*{\fill}
    \begin{subfigure}{0.48\textwidth}
        \centering
        \includegraphics[width=\linewidth]{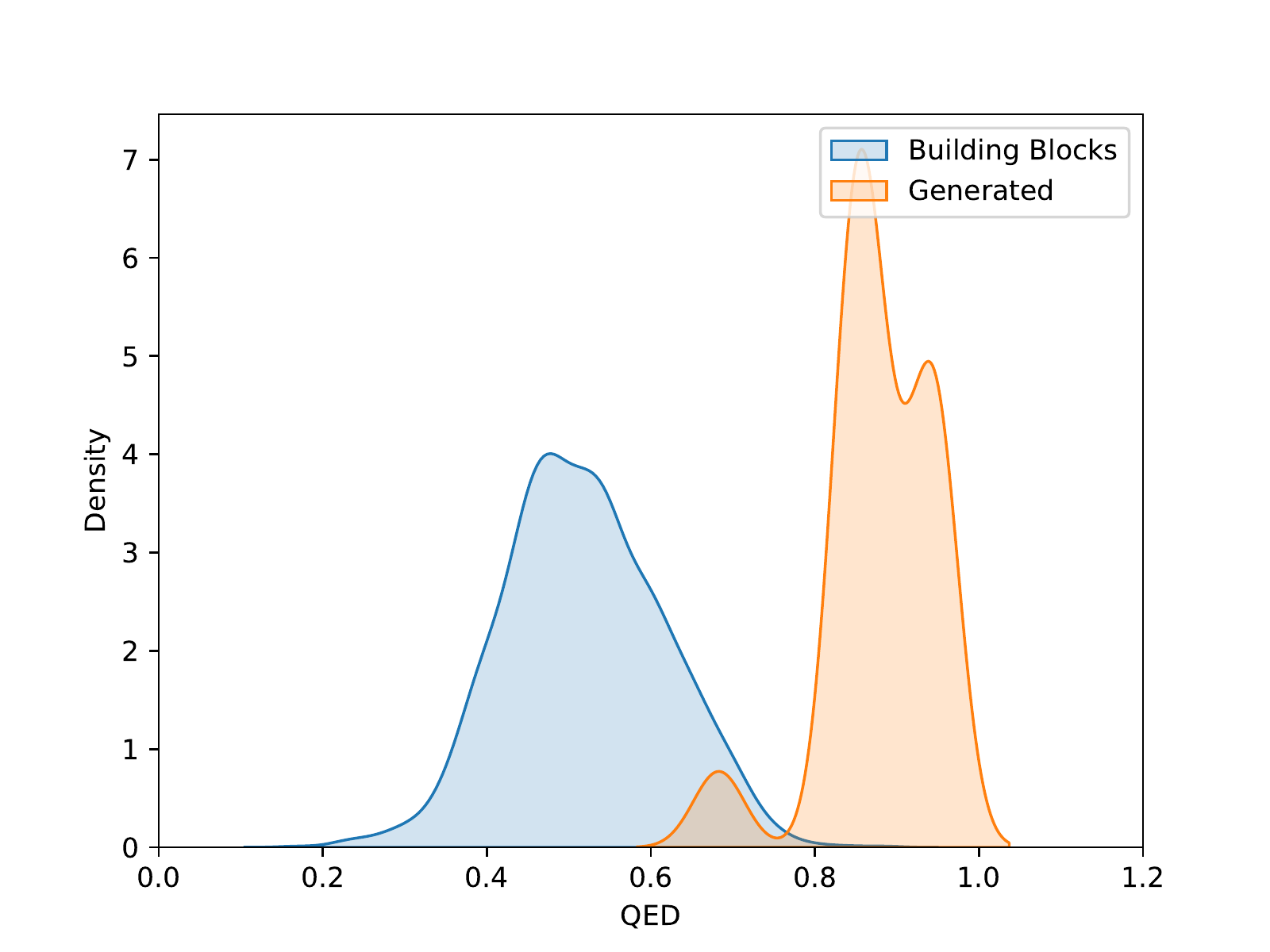}
        \caption{QED distribution of building blocks and generated molecules when QED is optimized.}
        \label{fig:dist_qed}        
    \end{subfigure}\\
    \begin{subfigure}{0.48\textwidth}
        \centering
        \includegraphics[width=\linewidth]{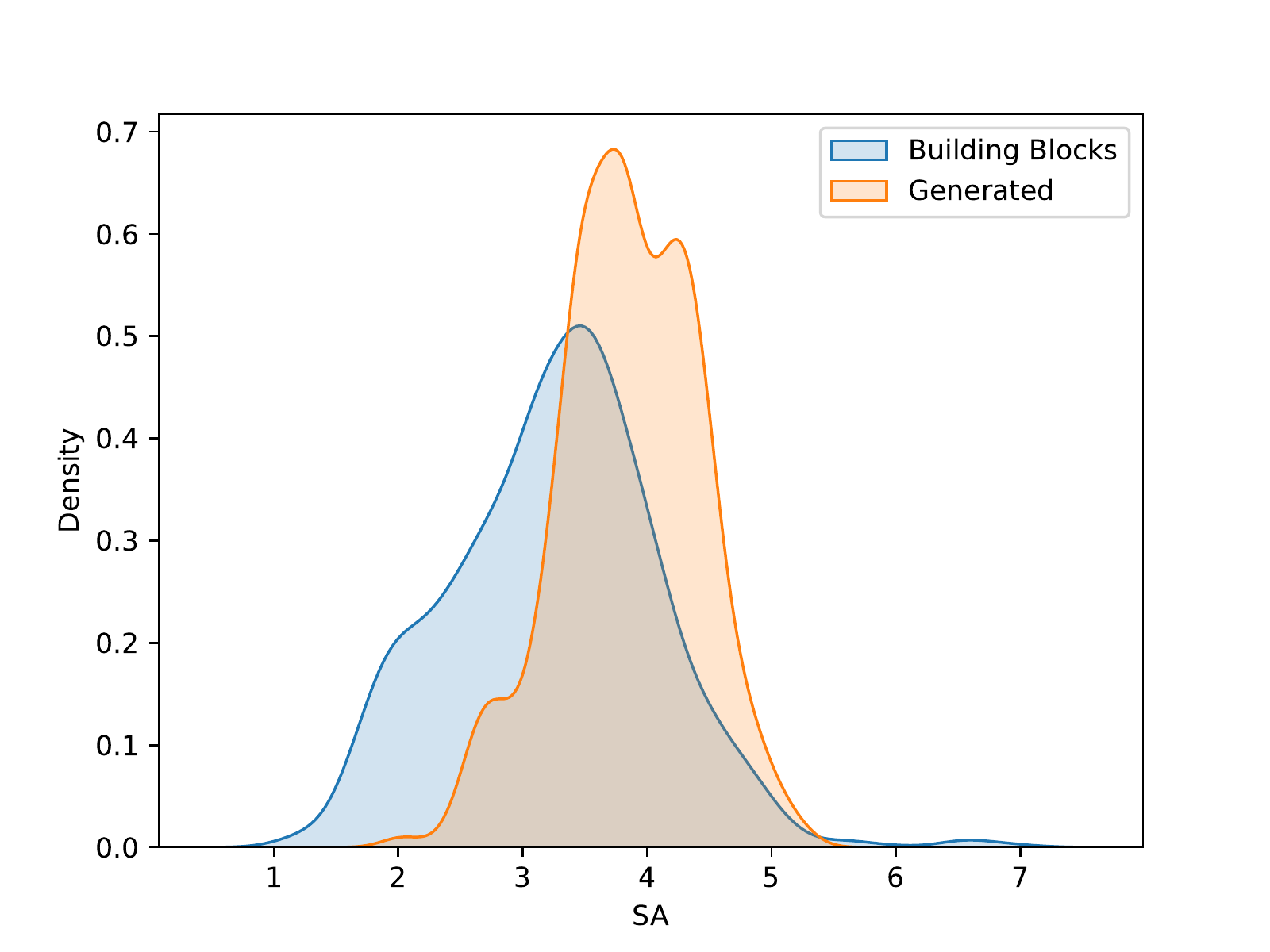}
        \caption{SA distribution of building blocks and generated molecules when both QED and SA are optimized simultaneously.}
        \label{fig:dist_sa_multi}        
    \end{subfigure}
    \hspace*{\fill}
    \begin{subfigure}{0.48\textwidth}
        \centering
        \includegraphics[width=\linewidth]{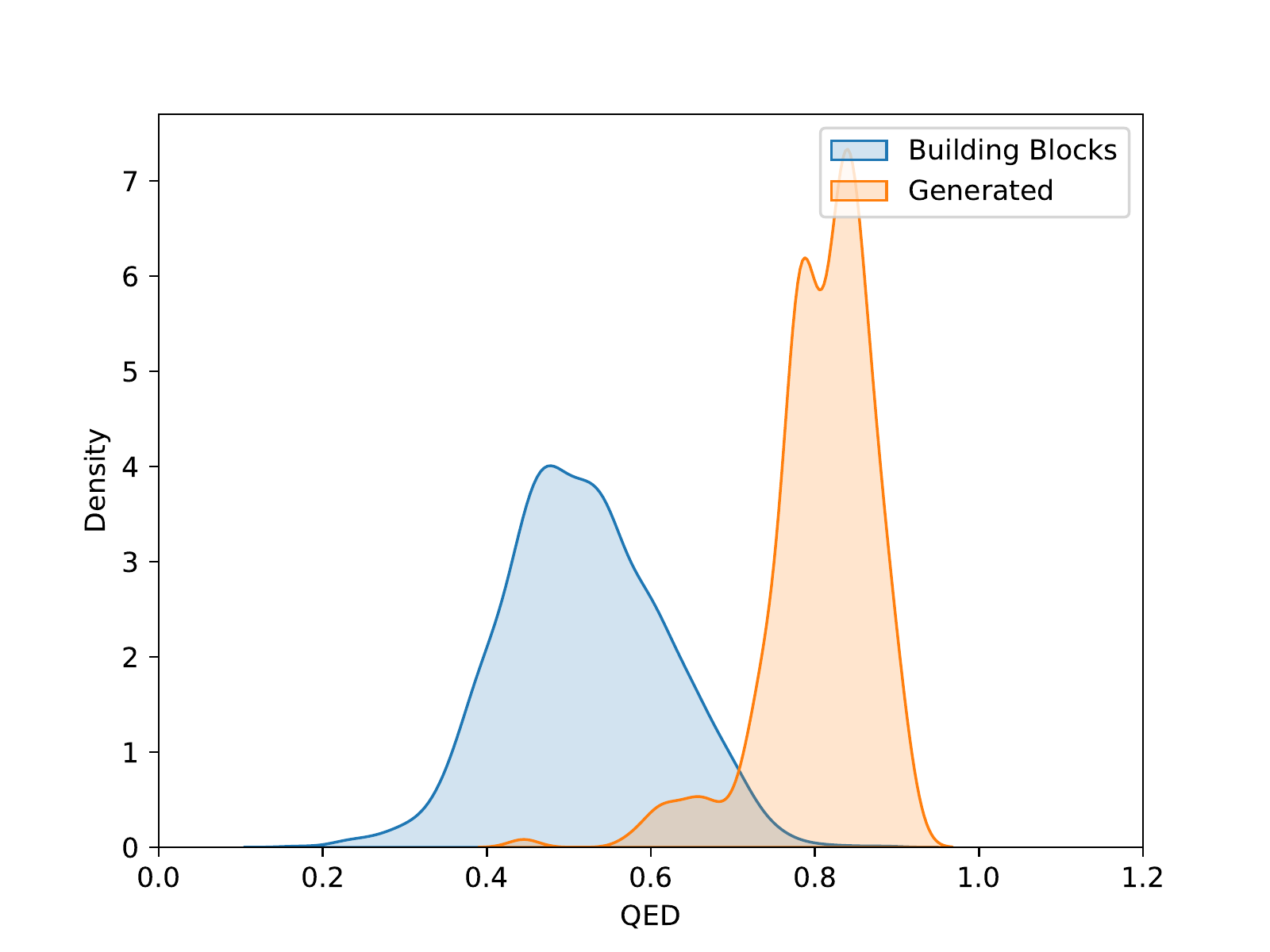}
        \caption{QED distribution of building blocks and generated molecules when both QED and SA are optimized simultaneously.}
        \label{fig:dist_qed_multi}        
    \end{subfigure}\\
    
    \caption{Distribution of generated compounds for optimizing both QED and SA in the multi-objective task on the GuacaMol dataset.}
    \label{fig:dist_qedsa} 
\end{figure}

In a multi-objective task with both QED and SA, Figure~\ref{fig:dist_qedsa} shows results of generated molecules when only QED is optimized or both QED and SA are optimized. If only QED is optimized (Figure~\ref{fig:dist_qedsa}(a), (b)), generated molecules tend to have a higher QED score with SA score being around 3 to 5. Otherwise if both QED and SA are optimized, the distribution of SA shifts left (Figure~\ref{fig:dist_qedsa}(c), (d)) compared to the case of QED optimization.  
 
\begin{table}[]
    \caption{Distribution benchmarks at 10k molecules on GuacaMol dataset.}
    \label{tab:distribution_benchmark}
    \centering
    \begin{tabular}{cccccc}
        \toprule
        benchmark & \begin{tabular}{c}Random\\Sampler\end{tabular} & \begin{tabular}{c}Graph\\MCTS\end{tabular}  &\begin{tabular}{c}SMILES\\LSTM\end{tabular} & VAE & \begin{tabular}{c}\textbf{MOLDR}\\(Random)\end{tabular} \\
        \midrule
         Validity & 1.00       &\textbf{1.000} & 0.959 & 0.870 & \textbf{1.000}\\
         Uniqueness & 0.997    & \textbf{1.000} & \textbf{1.000} & 0.999 & 0.994 \\
         Novelty  & 0.000      & 0.994 & 0.912 & 0.974 & \textbf{0.996 }\\
         KL divergence & 0.998 & 0.522 & \textbf{0.991} & 0.982  & 0.442 \\
         Fréchet ChemNet 
         Distance & 0.929 & 0.015 & \textbf{0.913} & 0.863 & 0.029 \\
         \bottomrule
    \end{tabular}
\end{table}

Results of distribution benchmarks are shown in Table~\ref{tab:distribution_benchmark}. This benchmark evaluates whether or not a model can generate valid, unique, and novel molecules from a training dataset. The KL (Kullback--Leibler) divergence and the Fréchet ChemNet distance (FCD) between a training set and generated molecules are also used. In a decomposition step, 1,709 building blocks are mined from the GuacaMol dataset with $\minsup = 10,000$. The distribution benchmark is evaluated from 10k sampled molecules in a reassembling step. MOLDR can generate valid molecules due to reassembling a building block of molecules. Although the uniqueness is slightly smaller than other models, MOLDR depends on random seeds to choose building blocks. In terms of the KL divergence and the FCD, MOLDR is inferior to SMILES LSTM and VAE. However, the score is similar to Graph MCTS because it is also a similar strategy to generate molecules. In addition, MOLDR can sample molecules randomly from an untrained policy network. Hence, MOLDR has a potential to generated molecules that are largely different from those in the training dataset, leading to lower scores of the KL divergence and the FCD. In order to improve the performance in terms of the KL divergence and the FCD, MOLDR would need to train the policy network and design an appropriate reward function, such as the similarity between the training dataset and generated molecules.  

\begin{figure}
\centering
    \begin{subfigure}{0.49\textwidth}
        \centering
        \includegraphics[width=\textwidth]{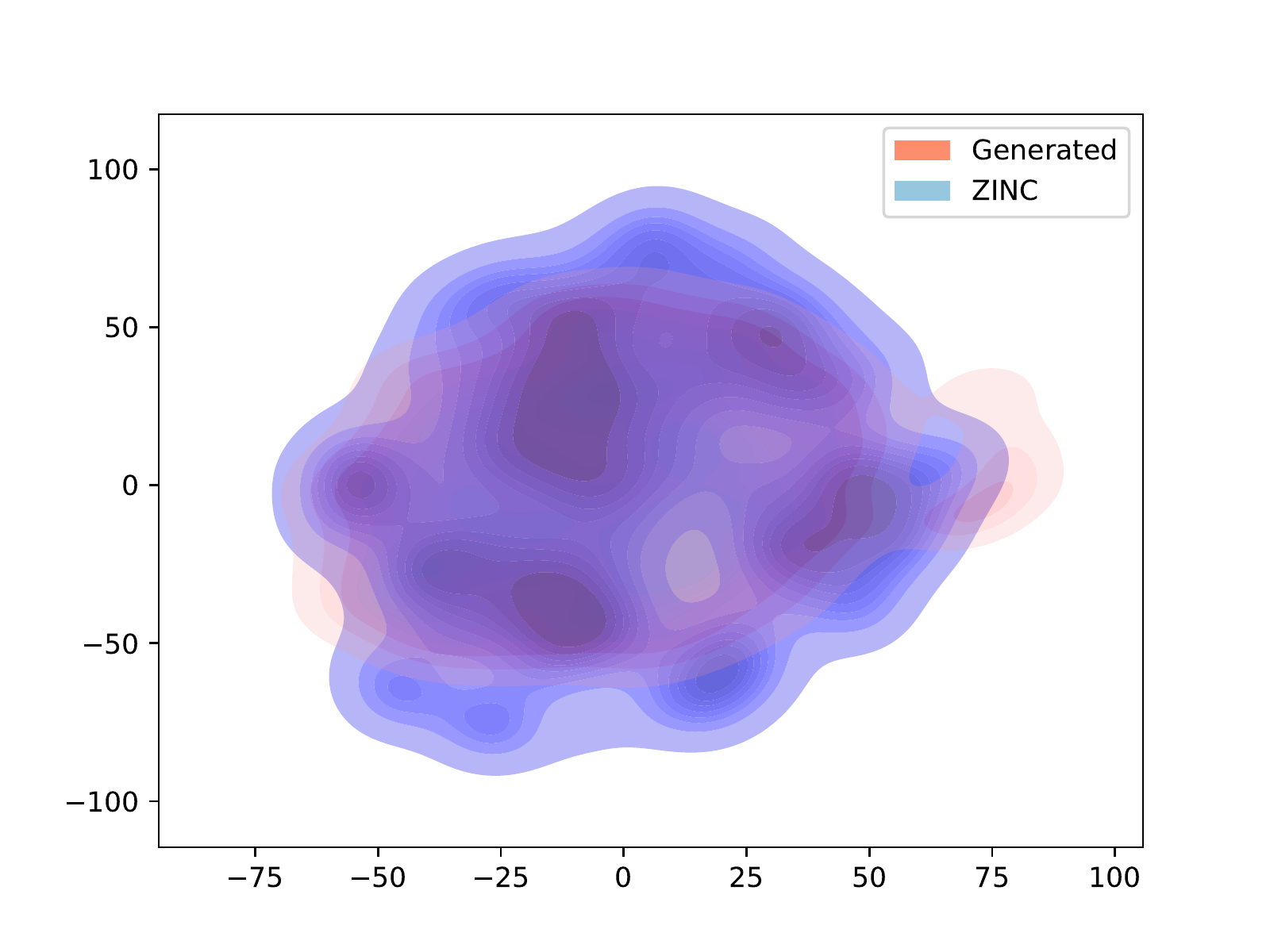}
        \caption{Distribution on ZINC dataset}
        \label{fig:dist_zinc}        
    \end{subfigure}    
    \begin{subfigure}{0.49\textwidth}
        \centering
        \includegraphics[width=\textwidth]{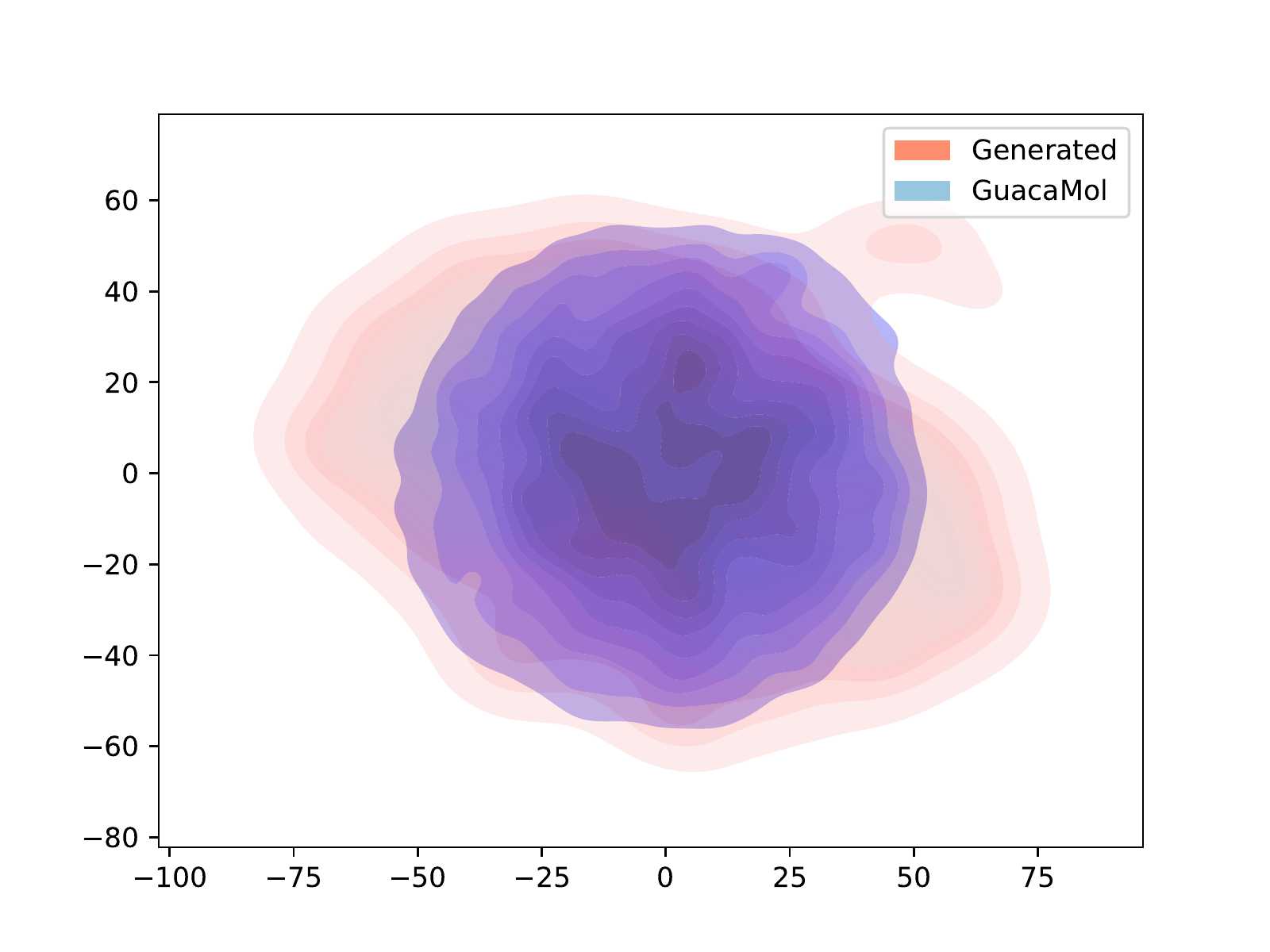}
        \caption{Distribution on GuacaMol dataset}
        \label{fig:dist_guacamol}        
    \end{subfigure}
    \caption{Distributions of training sets in GuacaMol and ZINC dataset, and generated molecules. Molecules are mapped into vectors using Mol2vec, and then we apply t-SNE dimensionality reduction for visualization.}
    \label{fig:dist_benchmark}
\end{figure}

Figure~\ref{fig:dist_benchmark} shows results of the distribution of generated molecules when trained on the ZINC or the GuacaMol dataset. Molecules are first mapped into 300-dimensional vectors using Mol2Vec and t-distributed stochastic neighbor embedding (t-SNE) is applied to visualize the distribution of generated molecules and that of the training set. On the ZINC dataset, generated molecules are mostly overlapped within the training set, while on the GuacaMol dataset, the distribution of generated molecules goes beyond that of the training set. It means that the generated molecules are not similar to the training set; therefore, the KL divergence and FCD scores are likely to be lower. However, random sampling from MOLDR can generate larger molecules out of distribution.

\begin{figure}
    \centering
    \includegraphics[width=\textwidth]{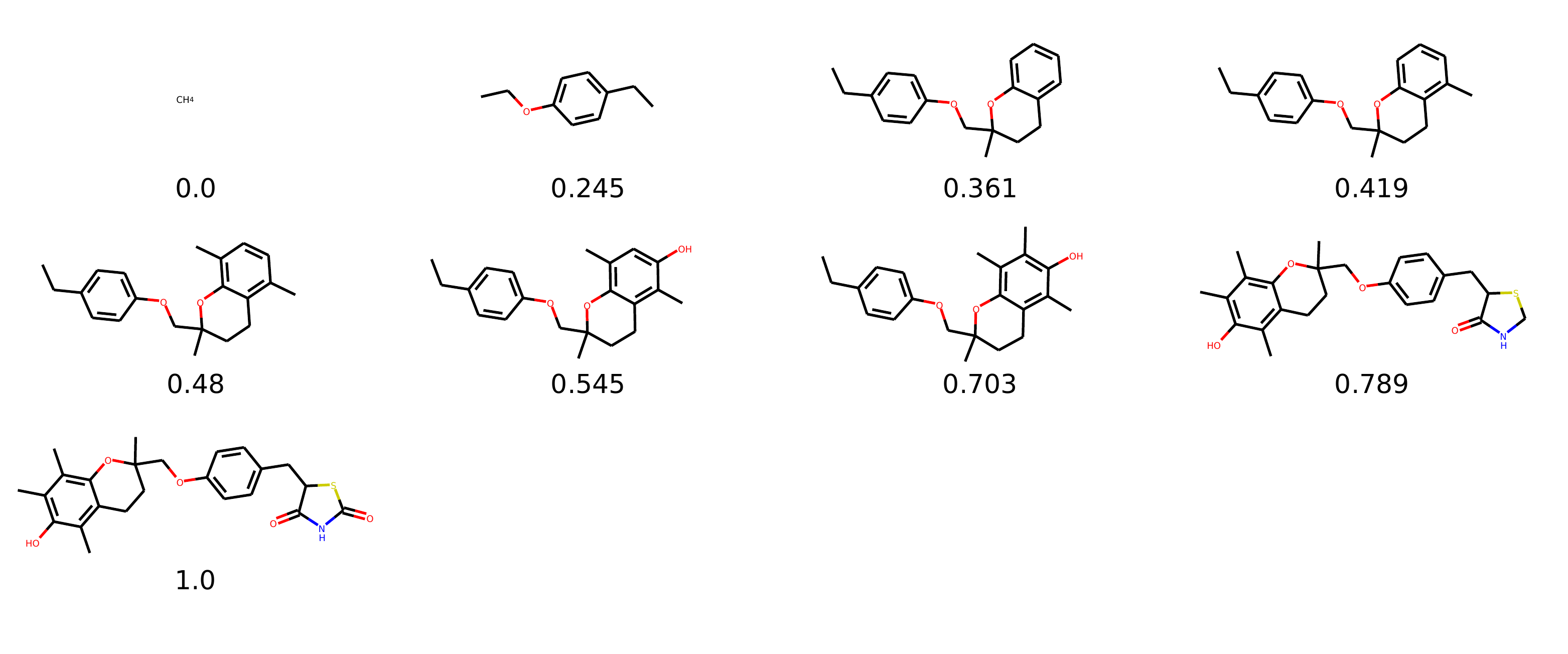}
    \caption{Generating process of Troglitazone rediscovery. The number under the molecules denotes the similarity score between a generated molecule and target. MOLDR can generate Troglitazone in 8 steps.}
    \label{fig:Troglitazone}
\end{figure}

\begin{figure}
    \centering
    \includegraphics[width=\textwidth]{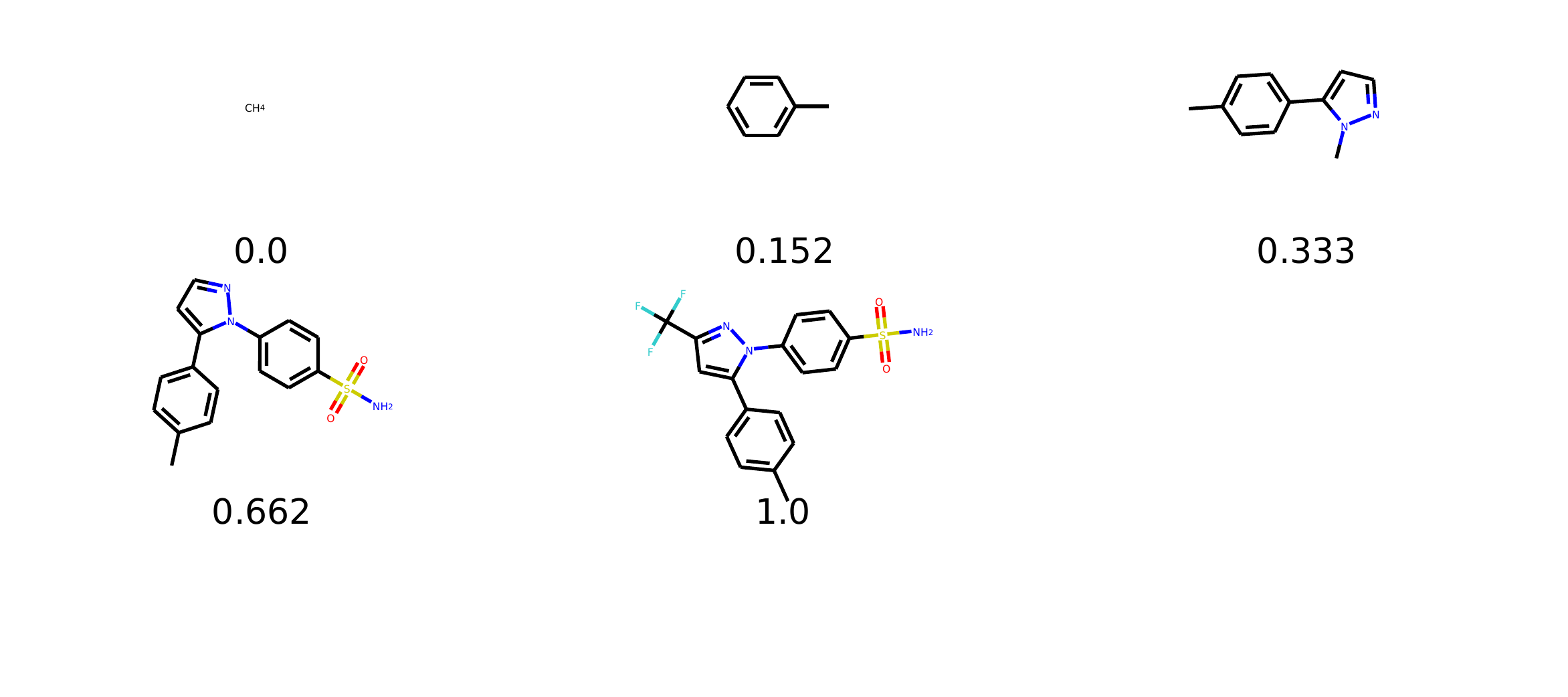}
    \caption{Generating process on Celecoxib rediscovery.}
    \label{fig:Celecoxib}
\end{figure}

Table~\ref{tab:guacamol_benchmark} shows the result of rediscovery benchmarks. MOLDR can generate the target molecules with high accuracy, whose scores are competitive with SMILES LSTM and Graph GA. The most notable difference between those models is that our model can visualize the generating process of molecules, not just generating the target molecule. Although in SMILES LSTM, intermediate molecules are evaluated from the state value function, and the SMILES character is selected based on the state, it is not easy to interpret why the character is vital at a particular time step, especially when generating a ring. In contrast, in MOLDR, building blocks are directly selected, and the sub-structure affects the target directly. Generating process is shown in Figure~\ref{fig:Celecoxib} and ~\ref{fig:Troglitazone}. Since the generation performance of MOLDR depends on the building blocks obtained from graph mining, in practical applications, it is important to prepare an appropriate dataset and set an appropriate minimum support based on a priori knowledge.

\begin{table*}[h]
    \centering
    \caption{Goal Directed benchmarks}
    \begin{tabular}{ccccc}
        \toprule
        Benchmark & Best in dataset &\begin{tabular}{c}SMILES\\LSTM\end{tabular} & Graph GA & MOLDR\\
        \midrule
        Celecoxib rediscovery & 0.505 & 1.000 &1.000 & 1.000 \\
        Troglitazone rediscovery & 0.419 & 1.000 &1.000 & 1.000  \\
        Aripiprazole similarity &  0.595 & 1.000 & 1.000 & 1.000 \\
        Osimertinib MPO & 0.839 & 0.907 & 0.953 & 0.898 \\ 
        Ranolazine MPO & 0.792  & 0.855 & 0.920 & 0.864 \\ 
        \bottomrule
    \end{tabular}
    \label{tab:guacamol_benchmark}
\end{table*}

\section{Conclusion}
We have proposed a new molecular generation method, called MOLDR, which decomposes graph structures and reassembles them. In our experiments on the ZINC database, MOLDR can find better molecules in terms of two properties, the penalized $\log P$ and the drug-likeness score QED, than the state-of-the-art molecular generation methods using Generative models and reinforcement learning. In terms of GuacaMol benchmarks, MOLDR can also reconstruct the target molecule if the substructures exist. Our approach is general, hence it can also be applied to any graph generation problem as well as molecular graph generation. MOLDR can also incorporate a priori knowledge about substructures by selecting specific datasets and/or designing reward functions.

In our future work, it is interesting to explore the clustering of subgraphs extracted by graph mining as there are often many similar graphs. Since the graph construction step is interpretable in MOLDR, incorporating MOLDR with the retro-synthesis analysis to consider the chemical reaction is an interesting topic.


\bibliography{references}

\providecommand{\latin}[1]{#1}
\makeatletter
\providecommand{\doi}
  {\begingroup\let\do\@makeother\dospecials
  \catcode`\{=1 \catcode`\}=2 \doi@aux}
\providecommand{\doi@aux}[1]{\endgroup\texttt{#1}}
\makeatother
\providecommand*\mcitethebibliography{\thebibliography}
\csname @ifundefined\endcsname{endmcitethebibliography}
  {\let\endmcitethebibliography\endthebibliography}{}
\begin{mcitethebibliography}{46}
\providecommand*\natexlab[1]{#1}
\providecommand*\mciteSetBstSublistMode[1]{}
\providecommand*\mciteSetBstMaxWidthForm[2]{}
\providecommand*\mciteBstWouldAddEndPuncttrue
  {\def\EndOfBibitem{\unskip.}}
\providecommand*\mciteBstWouldAddEndPunctfalse
  {\let\EndOfBibitem\relax}
\providecommand*\mciteSetBstMidEndSepPunct[3]{}
\providecommand*\mciteSetBstSublistLabelBeginEnd[3]{}
\providecommand*\EndOfBibitem{}
\mciteSetBstSublistMode{f}
\mciteSetBstMaxWidthForm{subitem}{(\alph{mcitesubitemcount})}
\mciteSetBstSublistLabelBeginEnd
  {\mcitemaxwidthsubitemform\space}
  {\relax}
  {\relax}

\bibitem[Polishchuk \latin{et~al.}(2013)Polishchuk, Madzhidov, and
  Varnek]{Polishchuk2013GDB17}
Polishchuk,~P.~G.; Madzhidov,~T.~I.; Varnek,~A. Estimation of the size of
  drug-like chemical space based on GDB-17 data. \emph{Journal of
  Computer-Aided Molecular Design} \textbf{2013}, \emph{27}, 675--679\relax
\mciteBstWouldAddEndPuncttrue
\mciteSetBstMidEndSepPunct{\mcitedefaultmidpunct}
{\mcitedefaultendpunct}{\mcitedefaultseppunct}\relax
\EndOfBibitem
\bibitem[Kirkpatrick and Ellis(2004)Kirkpatrick, and
  Ellis]{Kirkpatrick2004chemicalspace}
Kirkpatrick,~P.; Ellis,~C. Chemical space. \emph{Nature} \textbf{2004},
  \emph{432:823}\relax
\mciteBstWouldAddEndPuncttrue
\mciteSetBstMidEndSepPunct{\mcitedefaultmidpunct}
{\mcitedefaultendpunct}{\mcitedefaultseppunct}\relax
\EndOfBibitem
\bibitem[Kwon \latin{et~al.}(2019)Kwon, Bae, Jo, and Yoon]{QSAR}
Kwon,~S.; Bae,~H.; Jo,~J.; Yoon,~S. Comprehensive ensemble in QSAR prediction
  for drug discovery. \emph{BMC Bioinformatics} \textbf{2019}, \emph{20},
  521\relax
\mciteBstWouldAddEndPuncttrue
\mciteSetBstMidEndSepPunct{\mcitedefaultmidpunct}
{\mcitedefaultendpunct}{\mcitedefaultseppunct}\relax
\EndOfBibitem
\bibitem[Wong and Burkowski(2009)Wong, and Burkowski]{Wong2009IQSAR}
Wong,~W.~W.; Burkowski,~F.~J. A constructive approach for discovering new drug
  leads: Using a kernel methodology for the inverse-QSAR problem. \emph{Journal
  of Cheminformatics} \textbf{2009}, \emph{1}\relax
\mciteBstWouldAddEndPuncttrue
\mciteSetBstMidEndSepPunct{\mcitedefaultmidpunct}
{\mcitedefaultendpunct}{\mcitedefaultseppunct}\relax
\EndOfBibitem
\bibitem[Miyao \latin{et~al.}(2016)Miyao, Kaneko, and Funatsu]{Miyao2016IQSAR}
Miyao,~T.; Kaneko,~H.; Funatsu,~K. Inverse QSPR/QSAR Analysis for Chemical
  Structure Generation (from y to x). \emph{Journal of Chemical Information and
  Modeling} \textbf{2016}, \emph{56}, 286--299\relax
\mciteBstWouldAddEndPuncttrue
\mciteSetBstMidEndSepPunct{\mcitedefaultmidpunct}
{\mcitedefaultendpunct}{\mcitedefaultseppunct}\relax
\EndOfBibitem
\bibitem[Churchwell \latin{et~al.}(2004)Churchwell, Rintoul, Martin, Visco,
  Kotu, Larson, Sillerud, Brown, and Faulon]{Churchwell2004IQSAR}
Churchwell,~C.~J.; Rintoul,~M.~D.; Martin,~S.; Visco,~D.~P.; Kotu,~A.;
  Larson,~R.~S.; Sillerud,~L.~O.; Brown,~D.~C.; Faulon,~J.-L. The signature
  molecular descriptor: 3. Inverse-quantitative structure–activity
  relationship of ICAM-1 inhibitory peptides. \emph{Journal of Molecular
  Graphics and Modelling} \textbf{2004}, \emph{22}, 263--273\relax
\mciteBstWouldAddEndPuncttrue
\mciteSetBstMidEndSepPunct{\mcitedefaultmidpunct}
{\mcitedefaultendpunct}{\mcitedefaultseppunct}\relax
\EndOfBibitem
\bibitem[Gugisch \latin{et~al.}(2014)Gugisch, Kerber, Laue, Kohnert, Meringer,
  Rücker, and Wassermann]{Gugisch2014MOLGEN}
Gugisch,~R.; Kerber,~A.; Laue,~R.; Kohnert,~A.; Meringer,~M.; Rücker,~C.;
  Wassermann,~A. \emph{MOLGEN 5.0, A Molecular Structure Generator}; BENTHAM
  SCIENCE, 2014; pp 113--138\relax
\mciteBstWouldAddEndPuncttrue
\mciteSetBstMidEndSepPunct{\mcitedefaultmidpunct}
{\mcitedefaultendpunct}{\mcitedefaultseppunct}\relax
\EndOfBibitem
\bibitem[Takeda \latin{et~al.}(2020)Takeda, Hama, Hsu, Piunova, Zubarev,
  Sanders, Pitera, Kogoh, Hongo, Cheng, Bocanett, Nakashika, Fujita, Tsuchiya,
  Hino, Yano, Hirose, Toda, Orii, and Nakano]{Takeda2020MolIBM}
Takeda,~S. \latin{et~al.}  Molecular Inverse-Design Platform for Material
  Industries. Proceedings of the 26th ACM SIGKDD International Conference on
  Knowledge Discovery; Data Mining. 2020; p 2961–2969\relax
\mciteBstWouldAddEndPuncttrue
\mciteSetBstMidEndSepPunct{\mcitedefaultmidpunct}
{\mcitedefaultendpunct}{\mcitedefaultseppunct}\relax
\EndOfBibitem
\bibitem[Olivecrona \latin{et~al.}(2017)Olivecrona, Blaschke, Engkvist, and
  Chen]{Olivecrona2017REINVENT}
Olivecrona,~M.; Blaschke,~T.; Engkvist,~O.; Chen,~H. Molecular de-novo design
  through deep reinforcement learning. \emph{Journal of Cheminformatics}
  \textbf{2017}, \emph{9}\relax
\mciteBstWouldAddEndPuncttrue
\mciteSetBstMidEndSepPunct{\mcitedefaultmidpunct}
{\mcitedefaultendpunct}{\mcitedefaultseppunct}\relax
\EndOfBibitem
\bibitem[Gómez-Bombarelli \latin{et~al.}(2018)Gómez-Bombarelli, Wei,
  Duvenaud, Hernández-Lobato, Sánchez-Lengeling, Sheberla,
  Aguilera-Iparraguirre, Hirzel, Adams, and Aspuru-Guzik]{Rafa2018ChemVAE}
Gómez-Bombarelli,~R.; Wei,~J.~N.; Duvenaud,~D.; Hernández-Lobato,~J.~M.;
  Sánchez-Lengeling,~B.; Sheberla,~D.; Aguilera-Iparraguirre,~J.;
  Hirzel,~T.~D.; Adams,~R.~P.; Aspuru-Guzik,~A. Automatic Chemical Design Using
  a Data-Driven Continuous Representation of Molecules. \emph{ACS Central
  Science} \textbf{2018}, \emph{4}, 268–276\relax
\mciteBstWouldAddEndPuncttrue
\mciteSetBstMidEndSepPunct{\mcitedefaultmidpunct}
{\mcitedefaultendpunct}{\mcitedefaultseppunct}\relax
\EndOfBibitem
\bibitem[Kusner \latin{et~al.}(2017)Kusner, Paige, and
  Hernández-Lobato]{kusner2017grammar}
Kusner,~M.~J.; Paige,~B.; Hernández-Lobato,~J.~M. Grammar Variational
  Autoencoder. \emph{arXiv:1703.01925} \textbf{2017}, \relax
\mciteBstWouldAddEndPunctfalse
\mciteSetBstMidEndSepPunct{\mcitedefaultmidpunct}
{}{\mcitedefaultseppunct}\relax
\EndOfBibitem
\bibitem[Guimaraes \latin{et~al.}(2017)Guimaraes, Sanchez{-}Lengeling, Farias,
  and Aspuru{-}Guzik]{ORGAN}
Guimaraes,~G.~L.; Sanchez{-}Lengeling,~B.; Farias,~P. L.~C.; Aspuru{-}Guzik,~A.
  Objective-Reinforced Generative Adversarial Networks {(ORGAN)} for Sequence
  Generation Models. \emph{CoRR} \textbf{2017}, \emph{abs/1705.10843}\relax
\mciteBstWouldAddEndPuncttrue
\mciteSetBstMidEndSepPunct{\mcitedefaultmidpunct}
{\mcitedefaultendpunct}{\mcitedefaultseppunct}\relax
\EndOfBibitem
\bibitem[Jin \latin{et~al.}(2018)Jin, Barzilay, and Jaakkola]{Jin2018JTVAE}
Jin,~W.; Barzilay,~R.; Jaakkola,~T. Junction Tree Variational Autoencoder for
  Molecular Graph Generation. Proceedings of the 35th International Conference
  on Machine Learning, {ICML} 2018, Stockholmsm{\"{a}}ssan, Stockholm, Sweden,
  July 10-15, 2018. 2018; pp 2328--2337\relax
\mciteBstWouldAddEndPuncttrue
\mciteSetBstMidEndSepPunct{\mcitedefaultmidpunct}
{\mcitedefaultendpunct}{\mcitedefaultseppunct}\relax
\EndOfBibitem
\bibitem[Weininger(1988)]{Weininger1988SMILES}
Weininger,~D. {SMILES}, a chemical language and information system. 1.
  Introduction to methodology and encoding rules. \emph{Journal of Chemical
  Information and Computer Sciences} \textbf{1988}, \emph{28}, 31--36\relax
\mciteBstWouldAddEndPuncttrue
\mciteSetBstMidEndSepPunct{\mcitedefaultmidpunct}
{\mcitedefaultendpunct}{\mcitedefaultseppunct}\relax
\EndOfBibitem
\bibitem[You \latin{et~al.}(2018)You, Liu, Ying, Pande, and
  Leskovec]{You2018GCPN}
You,~J.; Liu,~B.; Ying,~Z.; Pande,~V.; Leskovec,~J. Graph Convolutional Policy
  Network for Goal-Directed Molecular Graph Generation. Proceedings of the 32nd
  International Conference on Neural Information Processing Systems. 2018; p
  6412–6422\relax
\mciteBstWouldAddEndPuncttrue
\mciteSetBstMidEndSepPunct{\mcitedefaultmidpunct}
{\mcitedefaultendpunct}{\mcitedefaultseppunct}\relax
\EndOfBibitem
\bibitem[Coulom(2006)]{MCTS_Go}
Coulom,~R. Efficient Selectivity and Backup Operators in Monte-Carlo Tree
  Search. Proceedings of the 5th International Conference on Computers and
  Games. Berlin, Heidelberg, 2006; p 72–83\relax
\mciteBstWouldAddEndPuncttrue
\mciteSetBstMidEndSepPunct{\mcitedefaultmidpunct}
{\mcitedefaultendpunct}{\mcitedefaultseppunct}\relax
\EndOfBibitem
\bibitem[Kocsis and Szepesv{\'a}ri(2006)Kocsis, and
  Szepesv{\'a}ri]{MCTS_bandit}
Kocsis,~L.; Szepesv{\'a}ri,~C. Bandit Based Monte-Carlo Planning. Machine
  Learning: ECML 2006. Berlin, Heidelberg, 2006; pp 282--293\relax
\mciteBstWouldAddEndPuncttrue
\mciteSetBstMidEndSepPunct{\mcitedefaultmidpunct}
{\mcitedefaultendpunct}{\mcitedefaultseppunct}\relax
\EndOfBibitem
\bibitem[Bickerton \latin{et~al.}(2012)Bickerton, Paolini, Besnard, Muresan,
  and Hopkins]{Bickerton2012QED}
Bickerton,~G.~R.; Paolini,~G.~V.; Besnard,~J.; Muresan,~S.; Hopkins,~A.~L.
  Quantifying the chemical beauty of drugs. \emph{Nature Chemistry}
  \textbf{2012}, \emph{4}, 90--98\relax
\mciteBstWouldAddEndPuncttrue
\mciteSetBstMidEndSepPunct{\mcitedefaultmidpunct}
{\mcitedefaultendpunct}{\mcitedefaultseppunct}\relax
\EndOfBibitem
\bibitem[Yang \latin{et~al.}(2017)Yang, Zhang, Yoshizoe, Terayama, and
  Tsuda]{Yang2017ChemTS}
Yang,~X.; Zhang,~J.; Yoshizoe,~K.; Terayama,~K.; Tsuda,~K. Chem{TS}: an
  efficient python library for de novo molecular generation. \emph{Science and
  Technology of Advanced Materials} \textbf{2017}, \emph{18}, 972--976\relax
\mciteBstWouldAddEndPuncttrue
\mciteSetBstMidEndSepPunct{\mcitedefaultmidpunct}
{\mcitedefaultendpunct}{\mcitedefaultseppunct}\relax
\EndOfBibitem
\bibitem[Yang \latin{et~al.}(2021)Yang, Aasawat, and
  Yoshizoe]{Yang2021MPChemTS}
Yang,~X.; Aasawat,~T.; Yoshizoe,~K. Practical Massively Parallel Monte-Carlo
  Tree Search Applied to Molecular Design. International Conference on Learning
  Representations. 2021\relax
\mciteBstWouldAddEndPuncttrue
\mciteSetBstMidEndSepPunct{\mcitedefaultmidpunct}
{\mcitedefaultendpunct}{\mcitedefaultseppunct}\relax
\EndOfBibitem
\bibitem[McKay(1998)]{MCKAY1998306}
McKay,~B.~D. Isomorph-Free Exhaustive Generation. \emph{Journal of Algorithms}
  \textbf{1998}, \emph{26}, 306--324\relax
\mciteBstWouldAddEndPuncttrue
\mciteSetBstMidEndSepPunct{\mcitedefaultmidpunct}
{\mcitedefaultendpunct}{\mcitedefaultseppunct}\relax
\EndOfBibitem
\bibitem[Stephen and Andrew(2009)Stephen, and Andrew]{MCKAY_Label}
Stephen,~H.~G.; Andrew,~R.~J. Mckay’s canonical graph labeling algorithm.
  \emph{In Communicating Mathematics} \textbf{2009}, \emph{479}, 99--111\relax
\mciteBstWouldAddEndPuncttrue
\mciteSetBstMidEndSepPunct{\mcitedefaultmidpunct}
{\mcitedefaultendpunct}{\mcitedefaultseppunct}\relax
\EndOfBibitem
\bibitem[Jin \latin{et~al.}(2020)Jin, Barzilay, and
  Jaakkola]{Jin2020multirationale}
Jin,~W.; Barzilay,~R.; Jaakkola,~T.~S. Composing Molecules with Multiple
  Property Constraints. \emph{CoRR} \textbf{2020}, \emph{abs/2002.03244}\relax
\mciteBstWouldAddEndPuncttrue
\mciteSetBstMidEndSepPunct{\mcitedefaultmidpunct}
{\mcitedefaultendpunct}{\mcitedefaultseppunct}\relax
\EndOfBibitem
\bibitem[Segler \latin{et~al.}(2018)Segler, Preuss, and Waller]{Segler2018}
Segler,~M. H.~S.; Preuss,~M.; Waller,~M.~P. Planning chemical syntheses with
  deep neural networks and symbolic AI. \emph{Nature} \textbf{2018},
  \emph{555}, 604--610\relax
\mciteBstWouldAddEndPuncttrue
\mciteSetBstMidEndSepPunct{\mcitedefaultmidpunct}
{\mcitedefaultendpunct}{\mcitedefaultseppunct}\relax
\EndOfBibitem
\bibitem[Zaki and Meira(2014)Zaki, and Meira]{Zaki2014}
Zaki,~M.~J.; Meira,~W.,~Jr. \emph{Data Mining and Analysis: Fundamental
  Concepts and Algorithms}; Cambridge University Press, 2014\relax
\mciteBstWouldAddEndPuncttrue
\mciteSetBstMidEndSepPunct{\mcitedefaultmidpunct}
{\mcitedefaultendpunct}{\mcitedefaultseppunct}\relax
\EndOfBibitem
\bibitem[Xifeng~Yan(2002)]{gSpan}
Xifeng~Yan,~J.~H. gSpan: Graph-Based Substructure Pattern Mining.
  \emph{International Conference on Data Mining} \textbf{2002}, 721--724\relax
\mciteBstWouldAddEndPuncttrue
\mciteSetBstMidEndSepPunct{\mcitedefaultmidpunct}
{\mcitedefaultendpunct}{\mcitedefaultseppunct}\relax
\EndOfBibitem
\bibitem[Houbraken \latin{et~al.}(2014)Houbraken, Demeyer, Michoel, Audenaert,
  Colle, and Pickavet]{ISMAGS}
Houbraken,~M.; Demeyer,~S.; Michoel,~T.; Audenaert,~P.; Colle,~D.; Pickavet,~M.
  The Index-Based Subgraph Matching Algorithm with General Symmetries (ISMAGS):
  Exploiting Symmetry for Faster Subgraph Enumeration. \emph{Plos One}
  \textbf{2014}, \emph{9}, 1--15\relax
\mciteBstWouldAddEndPuncttrue
\mciteSetBstMidEndSepPunct{\mcitedefaultmidpunct}
{\mcitedefaultendpunct}{\mcitedefaultseppunct}\relax
\EndOfBibitem
\bibitem[Segler \latin{et~al.}(2018)Segler, Kogej, Tyrchan, and
  Waller]{SMILES_RNN}
Segler,~M. H.~S.; Kogej,~T.; Tyrchan,~C.; Waller,~M.~P. Generating Focused
  Molecule Libraries for Drug Discovery with Recurrent Neural Networks.
  \emph{ACS central science} \textbf{2018}, \emph{4}, 120--131\relax
\mciteBstWouldAddEndPuncttrue
\mciteSetBstMidEndSepPunct{\mcitedefaultmidpunct}
{\mcitedefaultendpunct}{\mcitedefaultseppunct}\relax
\EndOfBibitem
\bibitem[Li \latin{et~al.}(2018)Li, Vinyals, Dyer, Pascanu, and
  Battaglia]{Li2018DGMG}
Li,~Y.; Vinyals,~O.; Dyer,~C.; Pascanu,~R.; Battaglia,~P. Learning Deep
  Generative Models of Graphs. 2018;
  \url{https://openreview.net/forum?id=Hy1d-ebAb}\relax
\mciteBstWouldAddEndPuncttrue
\mciteSetBstMidEndSepPunct{\mcitedefaultmidpunct}
{\mcitedefaultendpunct}{\mcitedefaultseppunct}\relax
\EndOfBibitem
\bibitem[Browne \latin{et~al.}(2012)Browne, Powley, Whitehouse, Lucas, Cowling,
  Rohlfshagen, Tavener, Perez, Samothrakis, and Colton]{MCTS_survey}
Browne,~C.~B.; Powley,~E.; Whitehouse,~D.; Lucas,~S.~M.; Cowling,~P.~I.;
  Rohlfshagen,~P.; Tavener,~S.; Perez,~D.; Samothrakis,~S.; Colton,~S. A Survey
  of Monte Carlo Tree Search Methods. \emph{IEEE Transactions on Computational
  Intelligence and AI in Games} \textbf{2012}, \emph{4}, 1--43\relax
\mciteBstWouldAddEndPuncttrue
\mciteSetBstMidEndSepPunct{\mcitedefaultmidpunct}
{\mcitedefaultendpunct}{\mcitedefaultseppunct}\relax
\EndOfBibitem
\bibitem[Silver \latin{et~al.}(2016)Silver, Huang, Maddison, Guez, Sifre,
  van~den Driessche, Schrittwieser, Antonoglou, Panneershelvam, Lanctot,
  Dieleman, Grewe, Nham, Kalchbrenner, Sutskever, Lillicrap, Leach,
  Kavukcuoglu, Graepel, and Hassabis]{Silver2016}
Silver,~D. \latin{et~al.}  Mastering the game of Go with deep neural networks
  and tree search. \emph{Nature} \textbf{2016}, \emph{529}, 484--489\relax
\mciteBstWouldAddEndPuncttrue
\mciteSetBstMidEndSepPunct{\mcitedefaultmidpunct}
{\mcitedefaultendpunct}{\mcitedefaultseppunct}\relax
\EndOfBibitem
\bibitem[Schulman \latin{et~al.}(2015)Schulman, Levine, Moritz, Jordan, and
  Abbeel]{Schulman2015TRPO}
Schulman,~J.; Levine,~S.; Moritz,~P.; Jordan,~M.~I.; Abbeel,~P. Trust Region
  Policy Optimization. 2015; \url{https://arxiv.org/abs/1502.05477}\relax
\mciteBstWouldAddEndPuncttrue
\mciteSetBstMidEndSepPunct{\mcitedefaultmidpunct}
{\mcitedefaultendpunct}{\mcitedefaultseppunct}\relax
\EndOfBibitem
\bibitem[Schulman \latin{et~al.}(2017)Schulman, Wolski, Dhariwal, Radford, and
  Klimov]{Schulman2017ppo}
Schulman,~J.; Wolski,~F.; Dhariwal,~P.; Radford,~A.; Klimov,~O. Proximal Policy
  Optimization Algorithms. 2017; \url{https://arxiv.org/abs/1707.06347}\relax
\mciteBstWouldAddEndPuncttrue
\mciteSetBstMidEndSepPunct{\mcitedefaultmidpunct}
{\mcitedefaultendpunct}{\mcitedefaultseppunct}\relax
\EndOfBibitem
\bibitem[Kakade and Langford(2002)Kakade, and Langford]{Kakade2002CPI}
Kakade,~S.; Langford,~J. Approximately Optimal Approximate Reinforcement
  Learning. Proceedings of the Nineteenth International Conference on Machine
  Learning. San Francisco, CA, USA, 2002; p 267–274\relax
\mciteBstWouldAddEndPuncttrue
\mciteSetBstMidEndSepPunct{\mcitedefaultmidpunct}
{\mcitedefaultendpunct}{\mcitedefaultseppunct}\relax
\EndOfBibitem
\bibitem[Jaeger \latin{et~al.}(2018)Jaeger, Fulle, and Turk]{Jaeger2018mol2vec}
Jaeger,~S.; Fulle,~S.; Turk,~S. Mol2vec: Unsupervised Machine Learning Approach
  with Chemical Intuition. \emph{Journal of Chemical Information and Modeling}
  \textbf{2018}, \emph{58}, 27--35\relax
\mciteBstWouldAddEndPuncttrue
\mciteSetBstMidEndSepPunct{\mcitedefaultmidpunct}
{\mcitedefaultendpunct}{\mcitedefaultseppunct}\relax
\EndOfBibitem
\bibitem[Mikolov \latin{et~al.}(2013)Mikolov, Sutskever, Chen, Corrado, and
  Dean]{Mikolov2013word2vec}
Mikolov,~T.; Sutskever,~I.; Chen,~K.; Corrado,~G.~S.; Dean,~J. Distributed
  Representations of Words and Phrases and their Compositionality. Advances in
  Neural Information Processing Systems. 2013\relax
\mciteBstWouldAddEndPuncttrue
\mciteSetBstMidEndSepPunct{\mcitedefaultmidpunct}
{\mcitedefaultendpunct}{\mcitedefaultseppunct}\relax
\EndOfBibitem
\bibitem[Brown \latin{et~al.}(2019)Brown, Fiscato, Segler, and
  Vaucher]{Brown2019GuacaMol}
Brown,~N.; Fiscato,~M.; Segler,~M.~H.; Vaucher,~A.~C. GuacaMol: Benchmarking
  Models for de Novo Molecular Design. \emph{Journal of Chemical Information
  and Modeling} \textbf{2019}, \emph{59}, 1096--1108\relax
\mciteBstWouldAddEndPuncttrue
\mciteSetBstMidEndSepPunct{\mcitedefaultmidpunct}
{\mcitedefaultendpunct}{\mcitedefaultseppunct}\relax
\EndOfBibitem
\bibitem[Liang \latin{et~al.}(2018)Liang, Liaw, Nishihara, Moritz, Fox,
  Goldberg, Gonzalez, Jordan, and Stoica]{Liang2018RLlib}
Liang,~E.; Liaw,~R.; Nishihara,~R.; Moritz,~P.; Fox,~R.; Goldberg,~K.;
  Gonzalez,~J.; Jordan,~M.; Stoica,~I. {RL}lib: Abstractions for Distributed
  Reinforcement Learning. Proceedings of the 35th International Conference on
  Machine Learning. 2018; pp 3053--3062\relax
\mciteBstWouldAddEndPuncttrue
\mciteSetBstMidEndSepPunct{\mcitedefaultmidpunct}
{\mcitedefaultendpunct}{\mcitedefaultseppunct}\relax
\EndOfBibitem
\bibitem[Paszke \latin{et~al.}(2019)Paszke, Gross, Massa, Lerer, Bradbury,
  Chanan, Killeen, Lin, Gimelshein, Antiga, Desmaison, Kopf, Yang, DeVito,
  Raison, Tejani, Chilamkurthy, Steiner, Fang, Bai, and Chintala]{PyTorch}
Paszke,~A. \latin{et~al.}  \emph{Advances in Neural Information Processing
  Systems 32}; Curran Associates, Inc., 2019; pp 8024--8035\relax
\mciteBstWouldAddEndPuncttrue
\mciteSetBstMidEndSepPunct{\mcitedefaultmidpunct}
{\mcitedefaultendpunct}{\mcitedefaultseppunct}\relax
\EndOfBibitem
\bibitem[Irwin \latin{et~al.}(2012)Irwin, Sterling, Mysinger, Bolstad, and
  Coleman]{Irwin2012ZINC}
Irwin,~J.~J.; Sterling,~T.; Mysinger,~M.~M.; Bolstad,~E.~S.; Coleman,~R.~G.
  ZINC: A Free Tool to Discover Chemistry for Biology. \emph{Journal of
  Chemical Information and Modeling} \textbf{2012}, \emph{52}, 1757--1768\relax
\mciteBstWouldAddEndPuncttrue
\mciteSetBstMidEndSepPunct{\mcitedefaultmidpunct}
{\mcitedefaultendpunct}{\mcitedefaultseppunct}\relax
\EndOfBibitem
\bibitem[Landrum(2006)]{RDKIT}
Landrum,~G. Rdkit: Open-source cheminformatics. \emph{Google Scholar}
  \textbf{2006}, \relax
\mciteBstWouldAddEndPunctfalse
\mciteSetBstMidEndSepPunct{\mcitedefaultmidpunct}
{}{\mcitedefaultseppunct}\relax
\EndOfBibitem
\bibitem[Ertl and Schuffenhauer(2009)Ertl, and Schuffenhauer]{Ertl2009SA}
Ertl,~P.; Schuffenhauer,~A. Estimation of synthetic accessibility score of
  drug-like molecules based on molecular complexity and fragment contributions.
  \emph{Journal of Cheminformatics} \textbf{2009}, \emph{1}, 8\relax
\mciteBstWouldAddEndPuncttrue
\mciteSetBstMidEndSepPunct{\mcitedefaultmidpunct}
{\mcitedefaultendpunct}{\mcitedefaultseppunct}\relax
\EndOfBibitem
\bibitem[Tan \latin{et~al.}(2022)Tan, Dai, Huang, Guo, Zheng, Lei, Chen, and
  Yang]{Tan2022DRlinker}
Tan,~Y.; Dai,~L.; Huang,~W.; Guo,~Y.; Zheng,~S.; Lei,~J.; Chen,~H.; Yang,~Y.
  DRlinker: Deep Reinforcement Learning for Optimization in Fragment Linking
  Design. \emph{Journal of Chemical Information and Modeling} \textbf{2022},
  \relax
\mciteBstWouldAddEndPunctfalse
\mciteSetBstMidEndSepPunct{\mcitedefaultmidpunct}
{}{\mcitedefaultseppunct}\relax
\EndOfBibitem
\bibitem[Preuer \latin{et~al.}(2018)Preuer, Renz, Unterthiner, Hochreiter, and
  Klambauer]{Preuer2018fcd}
Preuer,~K.; Renz,~P.; Unterthiner,~T.; Hochreiter,~S.; Klambauer,~G.
  Fr{\'e}chet ChemNet Distance: A Metric for Generative Models for Molecules in
  Drug Discovery. \emph{Journal of Chemical Information and Modeling}
  \textbf{2018}, \emph{58}, 1736--1741\relax
\mciteBstWouldAddEndPuncttrue
\mciteSetBstMidEndSepPunct{\mcitedefaultmidpunct}
{\mcitedefaultendpunct}{\mcitedefaultseppunct}\relax
\EndOfBibitem
\bibitem[Shi \latin{et~al.}(2020)Shi, Xu, Zhu, Zhang, Zhang, and
  Tang]{Chence2020graphaf}
Shi,~C.; Xu,~M.; Zhu,~Z.; Zhang,~W.; Zhang,~M.; Tang,~J. GraphAF: a Flow-based
  Autoregressive Model for Molecular Graph Generation. \textbf{2020}, \relax
\mciteBstWouldAddEndPunctfalse
\mciteSetBstMidEndSepPunct{\mcitedefaultmidpunct}
{}{\mcitedefaultseppunct}\relax
\EndOfBibitem
\end{mcitethebibliography}
\end{document}